\documentclass[a4paper]{article}

\usepackage{amssymb}
\usepackage{amsmath}
\usepackage{amsfonts}
\usepackage{graphicx}
\usepackage{fullpage}
\usepackage{authblk}
\usepackage{multirow}
\usepackage{pdflscape}
\usepackage{hyperref}
\usepackage{xcolor}
\renewcommand{\vec}{\mathbf}
\newcommand{\dct}[1]{\mathcal{C}_{#1}}
\newcommand{\dst}[1]{\mathcal{S}_{#1}}
\newcommand{\dft}{\mathcal{F}}
\newcommand{\idct}[1]{\mathcal{C}_{#1}^{-1}}
\newcommand{\idst}[1]{\mathcal{S}_{#1}^{-1}}
\newcommand{\idft}{\mathcal{F}^{-1}}
\newcommand{\kspace}{{\it k}-space}

\title{Pseudospectral time-domain (PSTD) methods for the wave equation: Realising boundary conditions with discrete sine and cosine transforms}

\author[1]{Elliott S. Wise}
\author[2]{Jiri Jaros}
\author[1]{Ben T. Cox}
\author[1,*]{Bradley E. Treeby}

\affil[1]{Department of Medical Physics and Biomedical Engineering, University College London, United Kingdom}
\affil[2]{Centre of Excellence IT4Innovations, Faculty of Information Technology, Brno University of Technology, Czech Republic \vspace{1em}}
\affil[*]{b.treeby@ucl.ac.uk}

\date{}

\begin{document}

\maketitle

\begin{abstract}
Pseudospectral time domain (PSTD) methods are widely used in many branches of acoustics for the numerical solution of the wave equation, including biomedical ultrasound and seismology. The use of the Fourier collocation spectral method in particular has many computational advantages. However, the use of a discrete Fourier basis is also inherently restricted to solving problems with periodic boundary conditions. Here, a family of spectral collocation methods based on the use of a sine or cosine basis is described. These retain the computational advantages of the Fourier collocation method but instead allow homogeneous Dirichlet (sound-soft) and Neumann (sound-hard) boundary conditions to be imposed. The basis function weights are computed numerically using the discrete sine and cosine transforms, which can be implemented using $O(N\log N)$ operations analogous to the fast Fourier transform. Practical details of how to implement spectral methods using discrete sine and cosine transforms are provided. The technique is then illustrated through the solution of the wave equation in a rectangular domain subject to different combinations of boundary conditions. The extension to boundaries with arbitrary real reflection coefficients or boundaries that are non-reflecting is also demonstrated using the weighted summation of the solutions with Dirichlet and Neumann boundary conditions.
\end{abstract}


\section{\label{sec_introduction}Introduction}
Simulating the propagation of acoustic waves requires the spatial and temporal derivative operators in the wave equation to be replaced with discrete formulas that can be implemented using a computer. A widely-used approach is the pseudospectral time domain (PSTD) method \cite{fornberg1994review,trefethen2000spectral}. This uses the Fourier collocation spectral method to calculate spatial gradients, and a finite-difference method to integrate forwards in time \cite{liu1999large}. For smooth wave fields, the error in the Fourier collocation spectral method decays exponentially with the number of grid nodes \cite{boyd2001chebyshev,hesthaven2007spectral}. This means only a small number of grid points per wavelength are required to reduce the effects of numerical dispersion, often close to the Nyquist limit of two points per wavelength (PPW) \cite{mast2001k,robertson2017accurate}. In comparison, the finite-difference time-domain (FDTD) method typically requires 10 to 20 PPW to minimise dispersion, and this number increases with the domain size due to the accumulation of numerical errors \cite{fornberg1987pseudospectral}. The difference in the number of PPW is critical for the tractability of many large-scale problems in which waves propagate across domains that are hundreds of wavelengths in each spatial dimension \cite{albin2011spectral,treeby2012modeling}.

\begin{figure}[t!]
    \centering
    \includegraphics[width=0.75\textwidth]{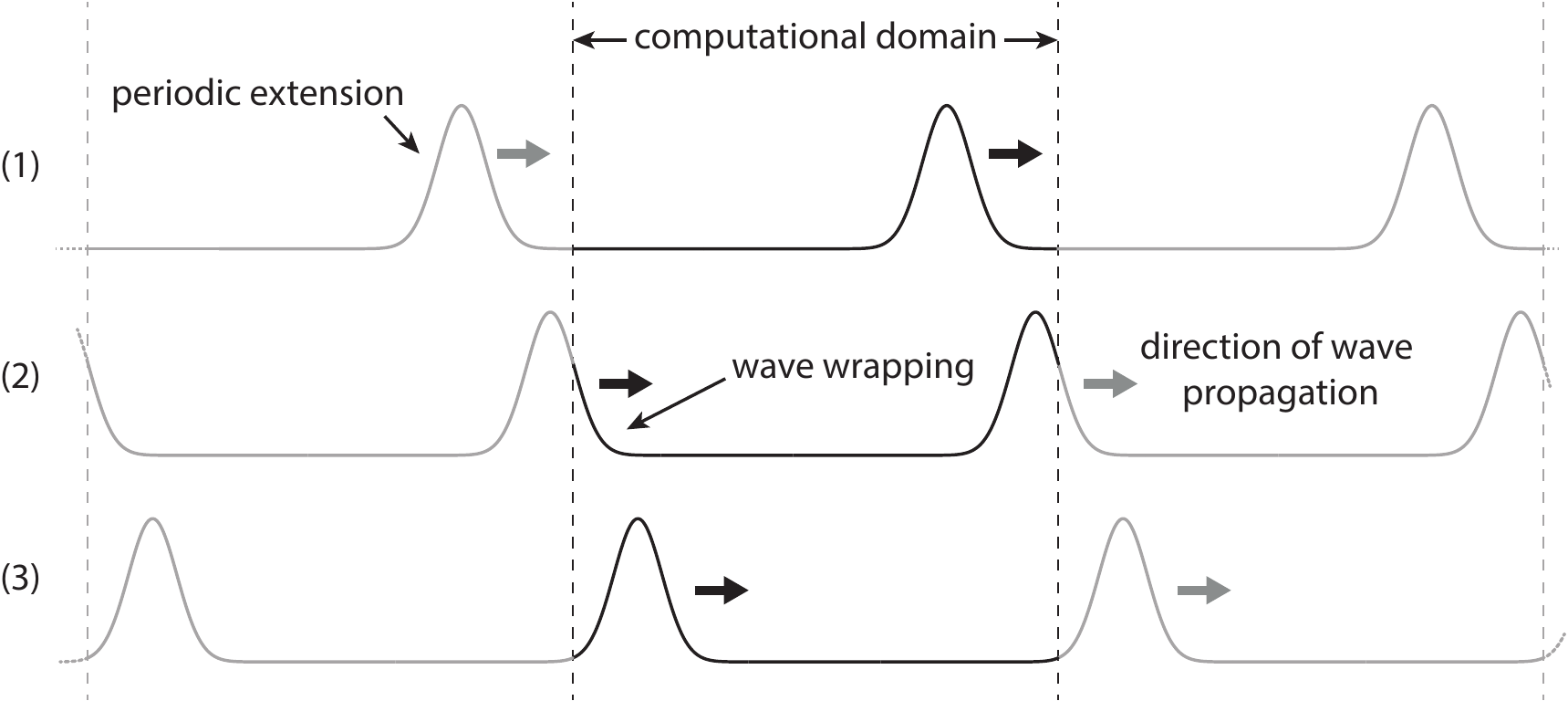}
    \caption{Wave wrapping due to the periodicity assumed by the discrete Fourier transform. Three snapshots in time are shown. As the wave exits the computational domain on the right, the wave from the periodically repeated image source appears on the left.}
    \label{fig_wave_wrapping}
\end{figure}

While the PSTD method has many advantages, the use of a discrete Fourier basis to compute spatial gradients is also inherently restricted to solving problems with periodic boundary conditions. This is because the discrete Fourier transform of a finite sequence assumes this sequence is periodically repeated outside its finite support. Practically, this means when the PSTD method is used to solve the wave equation, waves exiting the computational domain on one side will reappear on the other side. This can be understood in terms of image sources, where the pressure field is periodically (and infinitely) repeated as shown in Fig.\ \ref{fig_wave_wrapping}. In this example, as the wave within the computational domain exits to the right, the wavefield from the periodic image source appears within the domain on the left. In 1D where there is no decay of the wavefield with distance, this wave wrapping will continue \textit{ad infinitum}. 

In most circumstances in acoustics, a periodic boundary condition is neither physical nor desirable. To model non-reflecting boundaries (where the outgoing waves propagate as though in free-space), the periodicity is generally counteracted through the application of a perfectly matched layer (PML) \cite{berenger1994perfectly,liu1999large}. This is designed to absorb outgoing waves at the edge of the computational domain. A large body of literature now exists on the optimisation and implementation of PMLs and they can be highly effective \cite{berenger2007perfectly}. However, in some cases, other boundary conditions are required, and these are not straightforward to directly implement when using PSTD methods \cite{spa2010impedance,spa2011semi}.

To model reflecting boundaries, image sources can be used (assuming the wave propagation is linear). In this case, the true source and an image source are placed on opposites sides of the intended boundary (this idea was introduced by Lord Kelvin in the study of electrostatics \cite{jackson1999classical}). In the simplest case, the image source is placed equidistant from the intended boundary and is either  positive (in-phase) or negative (out-of-phase) relative to the true source, but otherwise identical. As the waves propagate, they pass each other at the (virtual) boundary and the summation of the two wave fields gives rise to the correct boundary and scattering behaviour. 

\begin{figure}[t!]
    \centering
    \includegraphics[width=0.8\textwidth]{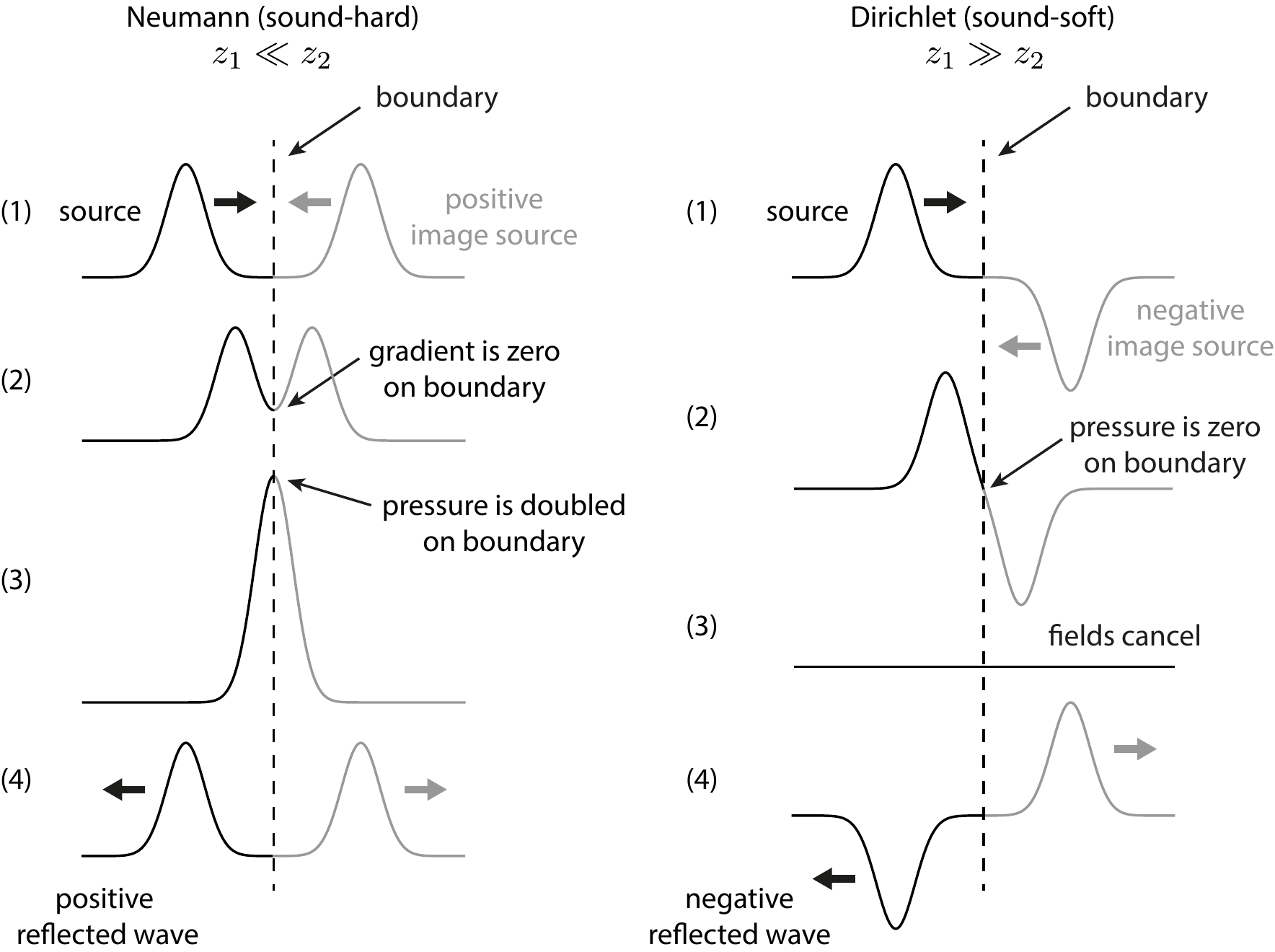}
    \caption{Neumann (sound-hard) and Dirichlet (sound-soft) boundary conditions realised using image sources. The Neumann case uses a positive (or in-phase) image source, which results in a pressure doubling and zero pressure gradient on the boundary. The Dirichlet case uses a negative (or out-of-phase) image source, which results in the pressure on the boundary always equal to zero. Acoustically, the Neumann case corresponds to the propagation of a wave from a low to high impedance medium (i.e., $z_1 \ll z_2$), where the reflected wave is in phase with the incident wave. The Dirichlet case corresponds to the propagation of a wave from a high to a low impedance medium (i.e., $z_1 \gg z_2$), where the reflected wave is out of phase with the incident wave. The numbers (1) to (4) correspond to four snapshots showing the interaction of the incident wave with the boundary.}
    \label{fig_image_sources}
\end{figure}

An illustrative example of using image sources to impose boundary conditions in 1D is given in Fig.\ \ref{fig_image_sources}. A positive image source corresponds to a sound-hard or Neumann boundary condition, where the gradient of the pressure field is always zero on the boundary, and the reflected wave is in phase with the incident wave. Considering the propagation of a wave between two acoustic half-spaces, this is equivalent to a boundary where the second medium has a much higher acoustic impedance. Similarly, a negative image source corresponds to a sound-soft or Dirichlet boundary condition, where the pressure is always zero on the boundary (this is sometimes called a pressure-release boundary \cite{kinsler1999fundamentals}), and the reflected wave is out-of-phase with the incident wave. Considering two acoustic half-spaces, this is equivalent to a boundary where the second medium has a much lower acoustic impedance. 

Using the idea of image sources, boundary conditions can be imposed within a PSTD wave model by increasing the size of the computational domain and directly including the image source terms. For example, this approach has been used to model wave propagation within reverberant cavities with reflecting walls \cite{cox2009photoacoustic}. However, the significant drawback is that the domain size must be increased in each Cartesian direction to allow the image sources to be defined. This can increase the computational load considerably, particularly in 3D. An alternative approach is to use another basis to compute the spatial gradients that already imposes the required boundary conditions. In the current work, the Fourier basis normally used in the PSTD method is replaced with a sine or cosine basis. These implicitly enforce homogeneous Dirichlet and Neumann boundary conditions while still retaining the same computational benefits as a Fourier collocation method \cite{strang1999discrete}.

Using sine and cosine transforms to compute spectral gradients as part of the PSTD method has previously been discussed by several authors. In acoustics, Kosloff \& Kosloff used a cosine basis to solve a one-dimensional wave equation with a Neumann (or sound-hard) boundary condition at each end of the domain \cite{kosloff1983non}. Sine and cosine transforms have also been used in elastodynamics for modelling applied forces that are symmetric or antisymmetric in the plane \cite{furumura19962}, and in electromagnetics to ensure field components vanish at the boundary, e.g., when modelling wave guides \cite{leung2001transformed,feise2004finite}. Outside of wave problems, sine and cosine transforms have also been used to impose Dirichlet and Neumann boundary conditions for reaction-diffusion problems \cite{bueno2006spectral}, to impose slip boundary conditions within fluid simulations \cite{long2009real}, to impose symmetric boundary conditions when solving the Elder problem in hydrology \cite{van2009insights}, to impose Dirichlet boundary conditions when solving the Klein-Gordon-Zakharov (KGZ) system \cite{bao2013exponential}, and to calculate spectral gradients through symmetric extensions \cite{ito2018pseudo}.

In the continuous case, the use of a cosine or sine basis to enforce Dirichlet or Neumann boundary conditions is straightforward to conceptualise---a sine wave is zero at the origin and thus satisfies a Dirichlet condition, while the gradient of a cosine wave is zero at the origin and thus satisfies a Neumann condition \cite{strang1999discrete}. However, in the discrete case, the different combinations of discrete symmetry leads to some complexity, with sixteen different discrete trigonometric transforms (DTTs) \cite{wang1985discretew,martucci1994symmetric} that each correspond to different boundary conditions and spatial grid definitions. To form a spectral collocation method, it is also important to understand how the discrete symmetry changes under differentiation and with the use of staggered computational grids.

In this paper, we review how the sixteen discrete trigonometric transforms fall into four distinct transform classes (see e.g., Ref.\ \cite{martucci1994symmetric}), and then discuss how each of these can be used to spectrally compute derivatives assuming different combinations of discrete symmetry. In \S2, the conventional Fourier PSTD method is reviewed for the wave equation. In \S3, the discrete cosine and sine transforms are reviewed, and the differentiation and interpolation properties of these transforms are discussed. An example of a DTT-based spectral collocation method is then given in \S4. In \S5, an extension to non-reflecting boundaries and arbitrary reflection coefficients is given. Discussion and summary are provided in \S6.


\section{The Fourier PSTD method for the wave equation}
Before discussing the machinery of using cosine and sine transforms to impose Dirichlet and Neumann boundary conditions, it is constructive to revisit the conventional Fourier PSTD method applied to the solution of the wave equation. The background to much of what follows can be found in several numerical methods texts, e.g.,  Refs.\ \cite{fornberg1998practical,trefethen2000spectral,boyd2001chebyshev,hesthaven2007spectral}. Consider the following coupled system of equations which describes linear acoustic wave propagation in a homogeneous medium in one-dimension 
\begin{align}
    \frac{\partial u}{\partial t} &= -\frac{1}{\rho_0}\frac{\partial p}{\partial x} \enspace, \nonumber \\
    \frac{\partial p}{\partial t} & = - \rho_0 c_0^2 \frac{\partial u}{\partial x} \enspace.
    \label{eq_wave_eq}
\end{align}
Here $u$ is the acoustic particle velocity, $p$ is the acoustic pressure, $\rho_0$ is the mass density, $c_0$ is the sound speed, and $x$ and $t$ represent the space and time variables, respectively. This system of coupled first-order equations is commonly used in computational acoustics instead of the second-order wave equation as it allows velocity inputs and outputs to be used, and a split-field perfectly matched layer to be applied \cite{tabei2002k}.

Now consider the numerical solution of Eq.\ \eqref{eq_wave_eq} using the Fourier PSTD method where $u$ is calculated on a temporally and spatially staggered grid relative to $p$, where the grid is shifted by half the temporal and spatial grid spacing (the use of a staggered grid improves the accuracy and stability of PSTD methods  \cite{fornberg1990high,chen1996staggered}). Three ingredients are required: (1) discretisation of the spatial gradients using the Fourier collocation spectral method, (2) transformation of the gradient values to and from the spatially staggered grid, and (3) discretisation of the temporal gradients using a finite difference scheme.

Starting with the spatial gradient calculation, in the Fourier collocation spectral method, the pressure and velocity fields are decomposed into a sum of complex exponential basis functions using a Fourier transform. In the continuous case, the forward and inverse Fourier transforms over one spatial dimension can be written as (other conventions can equally be used)
\begin{align}
    \dft\{f(x)\} \equiv g(k) &= \frac{1}{\sqrt{2\pi}} \int_{-\infty}^\infty f(x) e^{-ikx}\,dx\enspace, \nonumber \\
    \idft\{g(k)\} \equiv f(x) &= \frac{1}{\sqrt{2\pi}} \int_{-\infty}^\infty g(k) e^{ikx}\,dk\enspace,
    \label{eq_fourier_transform}
\end{align}
where $i=\sqrt{-1}$, $k$ is the spatial frequency or wavenumber, and $g(k)$ represents the complex basis function weights, i.e., the amplitude and phase of each frequency component needed to make up the original function $f(x)$. Using these transformations, it is straightforward to write an expression for the Fourier derivative of a bounded function, where \cite{trefethen2000spectral}
\begin{align}
    \frac{d}{dx}f(x) &= \frac{d}{dx} \frac{1}{\sqrt{2\pi}}  \int_{-\infty}^\infty g(k) e^{ikx}\,dk \nonumber \\
    &= \frac{1}{\sqrt{2\pi}}  \int_{-\infty}^\infty (ik) g(k) e^{ikx}\,dk \nonumber \\
    &= \idft\left\{ik\dft\left\{f(x)\right\} \right\}\enspace.
    \label{eq_fourier_diff}
\end{align}
In the discrete case, $f(x)$ is given by a finite periodic sequence, and the Fourier transforms are calculated using the discrete Fourier transform, generally using the fast Fourier transform (FFT). The discrete wavenumber vector $k$ is given by 
\begin{equation}
    k_n = \frac{2\pi}{N\Delta x}n\,\quad
    \mbox{where}\quad n = 
    \begin{cases}
        -\frac{N-1}{2},-\frac{N-1}{2}+1,\ldots,\frac{N-1}{2} & \mbox{if $N$ is odd} \\
        -\frac{N}{2},-\frac{N}{2}+1,\ldots,\frac{N}{2}-1 & \mbox{if $N$ is even}
    \end{cases} \enspace.
\end{equation}
Here, $N$ is the length of $f(x)$ (i.e., the number of samples in the sequence), and $\Delta x$ is the spacing between the grid points assuming the spatial samples are uniformly spaced. 

Next, to transform the calculated gradient values to and from the staggered grid, the shift property of the Fourier transform is used, where
\begin{equation}
\mathcal{F} \left\{ f(x + a) \right\} = e^{i k a} \mathcal{F} \left\{ f(x) \right\}\enspace.
\label{eq_fourier_shift_property}
\end{equation}
This follows from Eq.\ \eqref{eq_fourier_transform}. Here $a$ is a scalar shift variable which for a staggered grid is set to half the grid spacing, i.e., $\Delta x/2$.

Finally, to integrate forwards in time, an explicit first-order accurate forward difference is used \cite{strang2007computational}, where
\begin{equation}
    \frac{\partial f}{\partial t} \approx \frac{f^{n+1} - f^n}{\Delta t} \enspace.
    \label{eq_fd_time_integration}
\end{equation}
Here $f^n$ represents the discrete function values at the $n^\mathrm{th}$ time point, where $t = n \Delta t$ and $n = 0, 1, 2, \ldots$. Note, higher-order time integration schemes are also possible (e.g., Refs.\ \cite{wojcik1997pseudospectral,tabei2002k,lu2005time,hornikx2010extended,treeby2018nonstandard}). However, these are not discussed here as they do not influence the implementation of Neumann and Dirichlet boundary conditions using DTTs, which is the focus of the current work.

Combining the three ingredients then gives an explicit time-stepping solution to Eq.\ \eqref{eq_wave_eq} using the Fourier PSTD method on a staggered grid (solutions in higher dimensions follow analogously)
\begin{align}
    u^{n+\tfrac{1}{2}} &= u^{n-\tfrac{1}{2}} - \frac{\Delta t}{\rho_0} \idft\left\{ike^{i k \Delta x/2}\dft\left\{p^n\right\} \right\} \enspace, \nonumber \\
    p^{n+1} &= p^n - \Delta t \rho_0 c_0^2 \idft\left\{ike^{-i k \Delta x/2}\dft\left\{u^{n+\tfrac{1}{2}}\right\} \right\} \enspace.
    \label{eq_fourier_pstd_solution}
\end{align}
To run the model, initial values for $p$ and $u$ are required. Time dependent source terms can also be added \cite{jaros2016full}, but they are not discussed here. Note, the temporal staggering between the pressure and particle velocity arises because the update steps are interleaved with the spatial gradient calculations. As mentioned in \S1, the use of the discrete Fourier transform inherently assumes the wave field is periodically repeated, which leads to wave wrapping when the waves reach the edge of the computational domain (see Fig.\ \ref{fig_wave_wrapping}). In \S3, the Fourier transforms are replaced with sine and cosine transforms, and in \S4 these are applied to allow Dirichlet and Neumann boundary conditions to be imposed in the solution of the wave equation.


\section{Gradient calculations using discrete trigonometric transforms}


\subsection{The Fourier sine and cosine transforms}
Putting the numerical solution of the wave equation and the imposition of different boundary conditions to one side for a moment, consider now the Fourier sine and cosine transforms and their application to the calculation of spectral gradients. These transforms arise naturally as an extension of the Fourier transform (see e.g., Ref.\ \cite{britanak2007discrete}). This can be seen by rewriting the complex exponential basis functions of Eq.~\eqref{eq_fourier_transform} in terms of sines and cosines
\begin{equation}
    \dft\{f(x)\} = \frac{1}{\sqrt{2\pi}} \int_{-\infty}^\infty f(x) \left( \cos(kx) - i\sin(kx) \right)\,dx\,.
    \label{eq_fourier_transform_alt}
\end{equation}
If the function $f_o(x)$ is odd such that $f_o(-x)=-f_o(x)$, the cosine terms in Eq.~\eqref{eq_fourier_transform_alt} will integrate to zero, which leads to the definition of the forward and inverse Fourier sine transforms
\begin{align}
    \mathcal{S}\left\{f_o(x)\right\} &\equiv g_o(k) =  \sqrt{\frac{2}{\pi}}  \int_0^\infty f_o(x)\sin(kx)\,dx\enspace, \nonumber \\
    \mathcal{S}^{-1}\left\{g_o(k)\right\} &\equiv f_o(x) = \sqrt{\frac{2}{\pi}}  \int_0^\infty g_o(k)\sin(kx)\,dk\enspace,\enspace\enspace x \ge 0 \enspace.
\end{align}
If the input function $f_e(x)$ is instead even such that $f_e(-x)=f_e(x)$, the sine terms in Eq.~\eqref{eq_fourier_transform_alt} will integrate to zero, which leads to the definition of the forward and inverse Fourier cosine transforms
\begin{align}
    \mathcal{C}\left\{f_e(x)\right\} &\equiv g_e(k) = \sqrt{\frac{2}{\pi}} \int_0^\infty f_e(x)\cos(kx)\,dx\enspace, \nonumber \\
    \mathcal{C}^{-1}\left\{g_e(k)\right\} &\equiv f_e(x) = \sqrt{\frac{2}{\pi}} \int_0^\infty g_e(k)\cos(kx)\,dk\enspace,\enspace\enspace x \ge 0 \enspace.
\end{align}
Using these definitions, it is possible to show the derivative properties for the sine and cosine transforms. For an odd function this is
\begin{align}
    \frac{d}{dx}f_o(x)
    &= \frac{d}{dx} \left( \sqrt{\frac{2}{\pi}}  \int_0^\infty g_o(k)\sin(kx)\,dk \right) \enspace, \nonumber\\
    &= \sqrt{\frac{2}{\pi}} \int_0^\infty k g_o(k)\cos(kx)\,dk \enspace, \nonumber\\
    &= \mathcal{C}^{-1}\left\{k \mathcal{S}\left\{f_o(x)\right\} \right\}\enspace.
    \label{eq_FST_diff}
\end{align}
Similarly for an even function
\begin{equation}
    \frac{d}{dx}f(x)
    = \mathcal{S}^{-1}\left\{-k \mathcal{C}\left\{f_e(x)\right\} \right\}\,.
    \label{eq_FCT_diff}    
\end{equation}

Comparing Eqs.\ \eqref{eq_FST_diff} and \eqref{eq_FCT_diff} with Eq.~\eqref{eq_fourier_diff}, it can be seen that the differentiation properties for the cosine and sine transforms are very similar to those of the Fourier transform. The primary difference is that the function is multiplied by either $k$ or $-k$ after transformation to {\kspace} rather than by $-ik$, and the inverse transform differs from the forward transform. The latter reflects the fact that after differentiation, an odd function becomes even and vice versa. The cosine and sine transforms also produce real-valued outputs if the input function is real-valued, unlike the Fourier transform which produces a complex output.


\subsection{The discrete trigonometric transforms}

\begin{figure}[!th]
    \centering
    \includegraphics{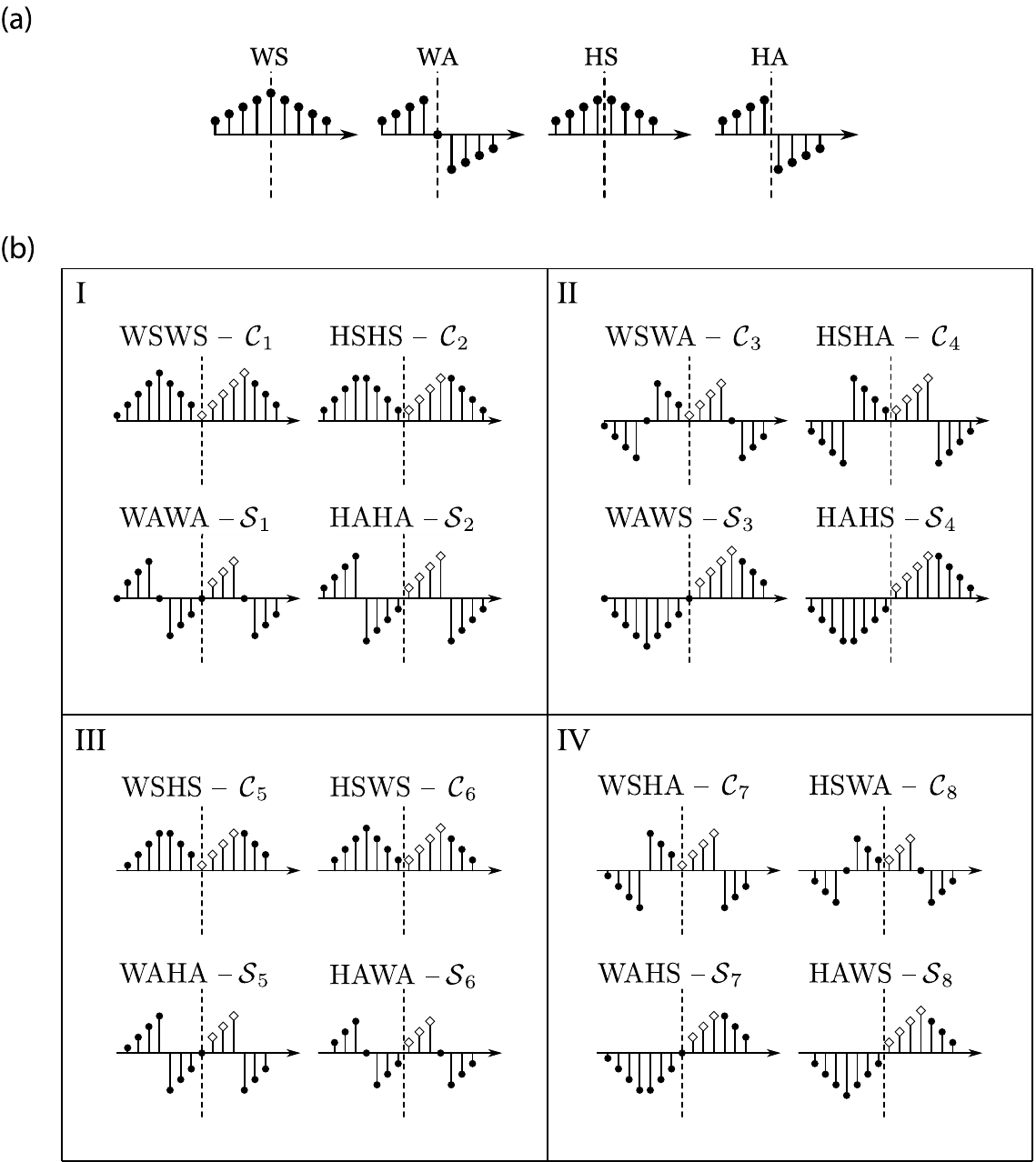}
    \caption{Discrete trigonometric transforms and symmetry (after Martucci). (a) The four types of discrete symmetry. The point of symmetry is indicated by the dashed line. (b) Examples of the 16 types of symmetric periodic sequence (SPS), and their corresponding DTTs.  The white diamonds indicate the representative samples, and the black circles give the implied symmetric-periodic extension.  Each SPS is either periodic or antiperiodic with period $M$.  For the example shown, $M=8$ for groups I and II, and $M=7$ for groups III and IV.}
    \label{fig_DTT_symmetries}
\end{figure}

In the continuous case, there are two possible symmetries a function may exhibit about $x=0$, even and odd. This leads to the definition of the two trigonometric transforms described in the previous section. However, in the discrete case, the number of possible symmetries increases to four \cite{wang1985discretew,martucci1994symmetric}. These symmetries are shown in Fig.~\ref{fig_DTT_symmetries}(a), and correspond to
\begin{itemize}
    \setlength\itemsep{0em}
    \item whole-sample symmetry (WS),
    \item whole-sample antisymmetry (WA),
    \item half-sample symmetry (HS), and 
    \item half-sample antisymmetry (HA).
\end{itemize}
The terms whole-sample (W) and half-sample (H) describe whether the point of symmetry lies on a grid point, or halfway between two grid points, respectively.
The terms symmetric (S) and antisymmetric (A) are analogous to even and odd in the continuous case. Together, the four types of symmetry may be used to define sixteen types of symmetric periodic sequence (SPS) by periodically extending each end of a discrete signal of finite length (called the representative sample) using one of the four symmetry types \cite{martucci1994symmetric}. The sixteen possible SPSs are shown in Fig.~\ref{fig_DTT_symmetries}(b). The naming convention is based on the symmetric extension used at each end. For instance, a WSHA-type sequence has a WS extension on its left end and a HA extension on its right. The extension is repeated indefinitely, such that the symmetry of a WSHA-type sequence is $\ldots$-WS-HA-WS-HA-$\dots$ and so on.

Each of the sixteen possible SPSs correspond to a unique discrete trigonometric transform (DTT) as given in Fig.~\ref{fig_DTT_symmetries}(b), where DST-I is denoted as $\dst{1}$ and DCT-I as $\dct{1}$, etc. The eight SPSs that have symmetric extensions on the left correspond to the eight types of discrete cosine transform (DCT), while the eight SPSs with antisymmetric extensions on the left correspond to the eight types of discrete sine transform (DST). The discrete equations for each of these transforms are defined in several places, including Refs.~\cite{martucci1994symmetric,britanak2007discrete,frigo2018fftw}, and are not repeated here. 

The set of sixteen DTTs (eight DCTs and eight DSTs) can be divided into four groups based on whether the corresponding SPS becomes periodic or antiperiodic after it has been symmetrically extended at one end, and whether this period $M$ is an even or odd number. Note here that a sequence is antiperiodic with period $M$ if $f(x) = -f(x+nM)$ for $n=1, 3, 5, \ldots$. Considering the first factor, groups I and III correspond to the SPSs that become periodic with period $M$, while groups II and IV are antiperiodic with period $M$. Considering the second factor, the implied period is always an even number for groups I and II, while the implied period is always an odd number for groups III and IV. The properties of these four groups are summarised in Table~\ref{table_DTT_wavenumber_groups} and Fig.~\ref{fig_DTT_symmetries}(b).

\begin{table}[t!]
    \renewcommand{\arraystretch}{1.4}
    \centering
    \caption{The sixteen DTTs and their corresponding symmetries grouped by wavenumber. The four groups of transforms correspond to those shown in Fig.~\ref{fig_DTT_symmetries}. Here $N$ is the length of the representative sample, $M$ is the corresponding implied period of the symmetric periodic sequence, $\Delta x$ is the grid spacing, and $k_n$ is the discrete wavenumber.}
    \label{table_DTT_wavenumber_groups}
    \resizebox{\textwidth}{!}{%
    \begin{tabular}{c c l l @{\hskip 1.5em}|@{\hskip 1.5em} c c l l}
        \hline
        \multicolumn{4}{c}{Periodic} 
        & \multicolumn{4}{c}{Antiperiodic}\\
        \multicolumn{4}{c}{$k_n=\frac{2\pi}{M\Delta x}n$} 
        & \multicolumn{4}{c}{$k_n=\frac{2\pi}{M\Delta x}\left(n+\frac{1}{2}\right)$} \\
        \hline
        Symmetry	& DTT			& $n=\ldots$ & $M=\ldots$ &
        Symmetry	& DTT			& $n=\ldots$ & $M=\ldots$\\
        \hline
        WSWS	& $\dct{1}$	& $0,1,\ldots,\frac{M}{2}$ & $2(N-1)$ &
        WSWA	& $\dct{3}$	& $0,1,\ldots,\frac{M}{2}-1$ & $2N$\\
        HSHS	& $\dct{2}$	& $0,1,\ldots,\frac{M}{2}-1$ & $2N$ &
        HSHA	& $\dct{4}$	& $0,1,\ldots,\frac{M}{2}-1$ & $2N$\\
        WAWA	& $\dst{1}$	& $1,2,\ldots,\frac{M}{2}-1$ & $2(N+1)$ &
        WAWS	& $\dst{3}$	& $0,1,\ldots,\frac{M}{2}-1$ & $2N$\\
        HAHA	& $\dst{2}$	& $1,2,\ldots,\frac{M}{2}$ & $2N$ &
        HAHS	& $\dst{4}$	& $0,1,\ldots,\frac{M}{2}-1$ & $2N$\\
        \hline
        WSHS	& $\dct{5}$	& $0,1,\ldots,\frac{M-1}{2}$ & $2N-1$&
        WSHA	& $\dct{7}$	& $0,1,\ldots,\frac{M-1}{2}$ & $2N-1$ \\
        HSWS	& $\dct{6}$	& $0,1,\ldots,\frac{M-1}{2}$ & $2N-1$ &
        HSWA	& $\dct{8}$	& $0,1,\ldots,\frac{M-1}{2}$ & $2N+1$ \\
        WAHA	& $\dst{5}$	& $1,2,\ldots,\frac{M+1}{2}$ & $2N+1$ &
        WAHS	& $\dst{7}$	& $0,1,\ldots,\frac{M-1}{2}$ & $2N+1$ \\
        HAWA	& $\dst{6}$	& $1,2,\ldots,\frac{M+1}{2}$ & $2N+1$ &
        HAWS	& $\dst{8}$	& $0,1,\ldots,\frac{M-1}{2}$ & $2N-1$\\
        \hline
    \end{tabular}}
\end{table}

When using DTTs to transform to and from $k$-space, attention must be given to the discrete wavenumber vector $k$, as this varies depending on the DTT used as shown in Table \ref{table_DTT_wavenumber_groups}. The different wavenumber ranges can be understood by considering the assumed symmetry of each sequence. For example, a sequence of WAWA-type is implicitly assumed to have a zero at either end of the representative sample (see Fig.~\ref{fig_DTT_symmetries}). Because of this, both the DC and Nyquist components of the signal must be zero. This is reflected in the range of wavenumbers for which the discrete trigonometric transform is defined, and thus the appropriate frequency components must be appended or removed when using combinations of transforms for which the wavenumber range varies. The different assumed symmetries between transforms also means the length of the sequence before and after differentiation may change, as the representative samples may be defined at different points in space.

Each of the DTTs is unitary (the basis vectors are orthogonal) and thus the inverse transformations can be defined in terms of the forward transforms as described in Refs.\ \cite{wang1985discretew,britanak2007discrete}. Letting $\mathcal{T}_j$ represent either $\dct{j}$ or $\dst{j}$, where $j=1, 2, \ldots, 8$ is the DCT or DST type, the inverse transform $\mathcal{T}_{j}^{-1}$ is given by
\begin{equation}
    \mathcal{T}_j^{-1} = \begin{cases}
        \frac{1}{M}\mathcal{T}_{j-1} & \mbox{if } j=3,7 \\
        \frac{1}{M}\mathcal{T}_j & \mbox{if } j=1,4,5,8 \\
        \frac{1}{M}\mathcal{T}_{j+1} & \mbox{if } j=2,6\,.
    \end{cases}
    \label{eq_inverse_transforms}
\end{equation}
The $1/M$ scaling factor comes from the discrete transform definitions used in the widely used FFTW library  in which the forward transforms are not normalised \cite{frigo2018fftw}. These relationships are practically useful when implementing DTT-based collocation methods. Note, however, it is the properties of the inverse transform (and not its related forward transform) that must be used when considering the range of wavenumbers and symmetry for which the transform is defined.


\subsection{Spectral gradient calculations}
Given a discrete input sequence (for example, the acoustic pressure or particle velocity defined on a regularly sampled Cartesian grid), spectral gradient calculation using a discrete trigonometric basis proceeds according to Eqs.~\eqref{eq_FST_diff} and \eqref{eq_FCT_diff}, where the continuous cosine and sine transforms are replaced by their discrete counterparts from Table~\ref{table_DTT_wavenumber_groups}. The forward transform is chosen based on the symmetry of the input sequence (which is linked to the desired boundary conditions as discussed in \S\ref{sec_introduction} and  \S\ref{dtt_pstd_formulation}). This then defines the appropriate inverse transform to use, as the symmetry of the function changes from symmetric to antisymmetric and vice versa after differentiation \cite{reeves2006shift}.

Consider an input sequence $f$ which has $N$ representative samples and WSWA symmetry. From Table \ref{table_DTT_wavenumber_groups}, the appropriate forward transform to use is $\dct{3}$. After differentiation, the symmetry of the output sequence will be WAWS, thus the corresponding inverse transform must be $\idst{3}$, which from Eq.\ \eqref{eq_inverse_transforms} can be computed using $\tfrac{1}{M}\dst{2}$. Using Eq.~\eqref{eq_FCT_diff}, the derivative can thus be computed via
\begin{equation}
    \frac{d}{dx}f = \idst{3}\{-k\dct{3}\{f\}\} = \frac{1}{M}\dst{2}\{-k\dct{3}\{f\}\} \enspace,
    \label{eq_dtt_gradient_example_I}
\end{equation}
where $M = 2N$. 

Despite the apparent simplicity of DTT--based gradient calculation, there is one complicating factor related to its numerical implementation. As mentioned in \S3.2, within each of the transform groups, the DTTs may be defined for different ranges of the same set of wavenumbers (see Table~\ref{table_DTT_wavenumber_groups}).
For instance, the $\dct{1}$ transform is defined for $n=0,1,\ldots,\frac{M}{2}$ where $k_n = \tfrac{2\pi}{M}n$, while the $\dst{1}$ transform is defined for $n=1,\ldots,\frac{M}{2}-1$. Thus, to differentiate a sequence of WSWS-type in an analogous manner to Eq.~\eqref{eq_dtt_gradient_example_I}, the first and last frequency components (corresponding to $n=0,\frac{M}{2}$) must be dropped from the discrete array before taking the inverse transform to obtain the differentiated result. For some combinations of DTTs, the reverse is true, and zeros must be appended and prepended to the discrete array before taking the inverse. The required sequence of operations for calculating the gradient of a function using DTTs is summarised in Table \ref{table_diff_periodic_sps} for periodic transforms (groups I and III) and in Table \ref{table_diff_antiperiodic_sps} for antiperiodic transforms (groups II and IV) in the Appendix.


\subsection{\label{sec:staggered}Interpolation on staggered grids}

DTTs can also be used for interpolation analogous to the shift property of the Fourier transform given in Eq.\ \eqref{eq_fourier_shift_property}. The general relations are discussed by Britanak \cite{britanak2007discrete}, although in some circumstances the expressions are rather complex. The simplest case is resampling a function from whole-sample symmetry to half-sample symmetry and vice-versa (this corresponds to shifting by half a grid point as required for staggered grid calculations). This can be achieved by converting to and from {\kspace} using the appropriate DTTs. For example, to interpolate a sequence of WSWA-type to a sequence of HSHA-type, the function is transformed into {\kspace} using $\dct{3}$, and back using $\idct{4}$, which from Eq.\ \eqref{eq_inverse_transforms} can be computed using $\tfrac{1}{M}\dct{4}$.
\begin{equation}
    g = \idct{4}\{\dct{3}\{f\}\} = \frac{1}{M}\dct{4}\{\dct{3}\{f\}\} \enspace.
\end{equation}
Again, attention must be paid to the range of wavenumbers for which the forward and inverse transforms are defined. The required sequence of operations for interpolating a function using DTTs is summarised in Table \ref{table_interp_periodic_sps} for periodic transforms (groups I and III) and in Table \ref{table_interp_antiperiodic_sps} for antiperiodic transforms (groups II and IV) in the Appendix.

\begin{figure}[!t]
    \centering
    \includegraphics{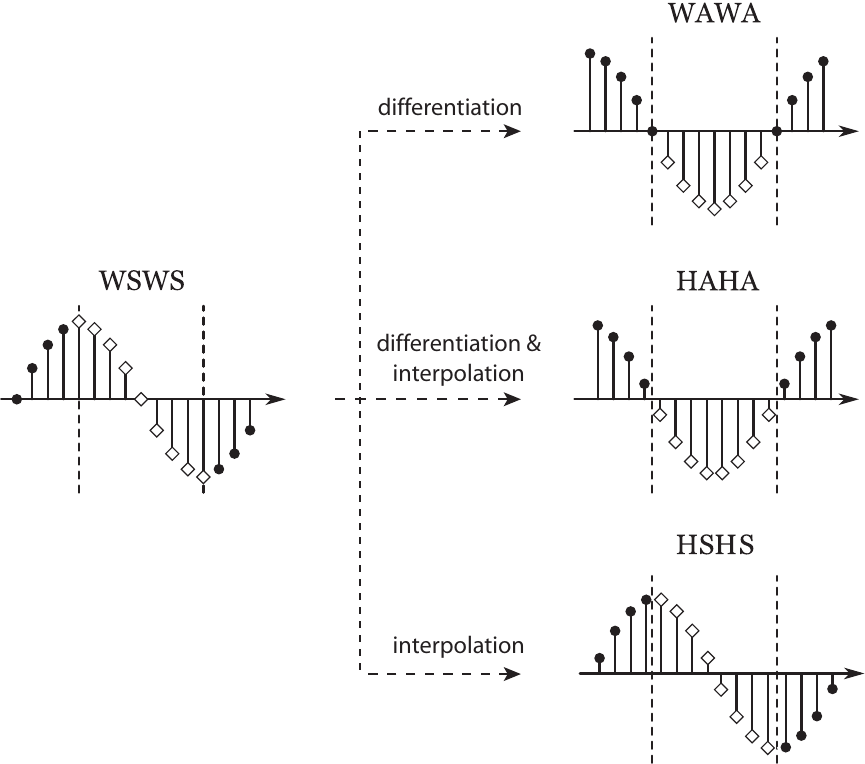}
    \caption{Illustrative example showing the transformation of a symmetric periodic sequence by differentiation, differentiation + interpolation onto a staggered grid, and interpolation onto a staggered grid. In each case, a different set of group I transformations is used which yields a different symmetry and representative sample size (illustrated with the white diamonds).}
    \label{figure_transform_example}
\end{figure}

Of particular interest in the context of PSTD methods is the combination of differentiation and interpolation, where the gradient values are resampled at points shifted forwards or backwards by half the grid point spacing \cite{fornberg1990high}. The combination of these operations follows analogously to the discussion of the individual differentiation and interpolation operations. The required sequence of operations for both differentiating and interpolating a function using DTTs is summarised in Table \ref{table_diff_interp_periodic_sps} for periodic transforms (groups I and III) and in Table \ref{table_diff_interp_antiperiodic_sps} for antiperiodic transforms (groups II and IV) in the Appendix. An illustration of a symmetric periodic sequence of WSWS-type after differentiation, interpolation, and differentiation and interpolation is shown in Fig.~\ref{figure_transform_example}.

Note, each of the four transform groups in Table~\ref{table_DTT_wavenumber_groups} is closed under differentiation and interpolation. Because of the direct relationship between DTTs and SPSs, this is equivalent to saying that differentiating or interpolating a SPS will always produce another SPS in the same group. The closed nature of these groups is clear when considering that under differentiation the symmetry of a function changes (symmetric $\leftrightarrow$ antisymmetric) and that under interpolation the point of symmetry changes (whole-sample $\leftrightarrow$ half-sample).


\section{The PSTD method with discrete trigonometric transforms}


\subsection{\label{dtt_pstd_formulation}Formulation}
Now that the required ingredients to form a PSTD method using DTTs have been introduced, we can return to the solution of the wave equation given in Eq.\ \eqref{eq_wave_eq} subject to different boundary conditions at each end of the domain. In 1D, there are four possible combinations of the left-right boundary condition: Neumann-Neumann, Neumann-Dirichlet, Dirichlet-Neumann, and Dirichlet-Dirichlet. Assuming the acoustic pressure $p$ is calculated on the regular grid and the particle velocity $u$ on the staggered grid (analogous to the Fourier PSTD solution given in Eq.\ \eqref{eq_fourier_pstd_solution}), these correspond to symmetries for the acoustic pressure field of WSWS, WSWA, WAWS, and WAWA, respectively. The particle velocity will instead have half-sample symmetries (i.e., W$\rightarrow$H) with the symmetries reversed (i.e., S$\rightarrow$A and A$\rightarrow$S). These symmetries are summarised in Table \ref{table_transform_choices}.

Given the chosen boundary conditions at each end of the domain (and thus the appropriate symmetries and DTTs to use), the spatial gradients shifted to and from the staggered grid can be calculated as discussed in \S\ref{sec:staggered}. Using the first-order accurate forward difference from Eq.\ \eqref{eq_fd_time_integration} to integrate in time, a DTT-based PSTD solution for Eq.\ \eqref{eq_wave_eq} can be written as
\begin{subequations}
    \begin{align}
        u^{n+\tfrac{1}{2}} &= u^{n-\tfrac{1}{2}} - \frac{\Delta t}{\rho_0} \mathcal{T}_2^{-1}\left\{a_1\mathcal{T}_1\{p^n\}\right\} \enspace, \label{eq_dtt_pstd_solution_a} \\
        p^{n+1} &= p^n - \Delta t \rho_0 c_0^2 \mathcal{T}_1^{-1}\left\{a_2\mathcal{T}_2\left\{u^{n+\frac{1}{2}}\right\}\right\} \enspace. \label{eq_dtt_pstd_solution_b}
    \end{align}
    \label{eq_dtt_pstd_solution}
\end{subequations}
\noindent The transforms $\mathcal{T}_1$ and $\mathcal{T}_2$ and {\kspace} coefficients $a_1$ and $a_2$ for the different boundary conditions are given in Table~\ref{table_transform_choices}. As $u$ has the same symmetry as $\partial p /\partial x$, the forward transform in Eq.\ \eqref{eq_dtt_pstd_solution_b} matches the inverse transform used in Eq.\ \eqref{eq_dtt_pstd_solution_a} and vice-versa.

\begin{table}[h]
    \renewcommand{\arraystretch}{1.3}
    \centering
    \caption{Transforms and {\kspace} coefficients used for the DTT-based PSTD solution of the wave equation given in Eq.~\eqref{eq_dtt_pstd_solution}. The {\kspace} coefficients in the periodic case also include shift terms to move to and from the staggered grid points. These shifts are implicit in the DTT pairs specified.}
    \label{table_transform_choices}
        \begin{tabular}{c c c c c c c c}
        \hline
        \multicolumn{2}{c}{Boundary condition} & \multicolumn{2}{c}{Sequence type} & \multicolumn{2}{c}{Transforms} & \multicolumn{2}{c}{{\kspace} coefficient}\\
        left & right & $p$ & $u$ & $\mathcal{T}_1$ & $\mathcal{T}_2$ & $a_1$ & $a_2$\\
        \hline
        Neumann & Neumann & WSWS & HAHA & $\dct{1}$ & $\dst{2}$ & $-k$ & $k$ \\
        Neumann & Dirichlet & WSWA & HAHS & $\dct{3}$ & $\dst{4}$ & $-k$ & $k$ \\
        Dirichlet & Neumann & WAWS & HSHA & $\dst{3}$ & $\dct{4}$ & $k$ & $-k$ \\
        Dirichlet & Dirichlet & WAWA & HSHS & $\dst{1}$ & $\dct{2}$ & $k$ & $-k$ \\
        \hline
        \multicolumn{4}{c}{Periodic} & $\dft$ & $\dft$ & $ike^{ik\frac{\Delta x}{2}}$ & $ik e^{-ik\frac{\Delta x}{2}}$ \\
        \hline
    \end{tabular}
\end{table}


\subsection{\label{sec_numerical_results}Numerical results}
The discrete equations given in Eq.~\eqref{eq_dtt_pstd_solution} were implemented in MATLAB (R2019b), with the DTTs computed using a MEX interface to the C++ FFTW library \cite{frigo2005design}. FFTW provides an efficient implementation of both the FFT and the group I and II DTTs \cite{frigo2018fftw}. MATLAB uses FFTW to compute FFTs as part of its inbuilt functions, however, it does not currently provide functions that link to the DTT implementations in FFTW. The MEX interface and example scripts used in this paper are available open-source from GitHub \cite{treeby2020dttlibrary}.

To demonstrate the effect of the different boundary conditions on the propagation of an acoustic wave, five simulations were performed using each of the transform combinations outlined in Table~\ref{table_transform_choices}. The grid size was set to $N = 256$, with a grid spacing of $\Delta x = 1/N$. The sound speed and density were set to the properties of water, with $c_0 = 1500$ m.s$^{-1}$ and $\rho_0 = 1000$ kg.m$^{-3}$, and the time step was given by a Courant-Friedrichs-Lewy (CFL) number of 0.2, where $\Delta t = \text{CFL} \Delta x / c_0$. The initial pressure distribution $p_0$ was defined as a Gaussian, and the initial particle velocity $u_0$ was defined as $u_0 = -p_0 / (\rho_0 c_0)$. This gives a travelling wave propagating in the negative x-direction.

\begin{figure}[!ptbh]
    \includegraphics[width=\textwidth]{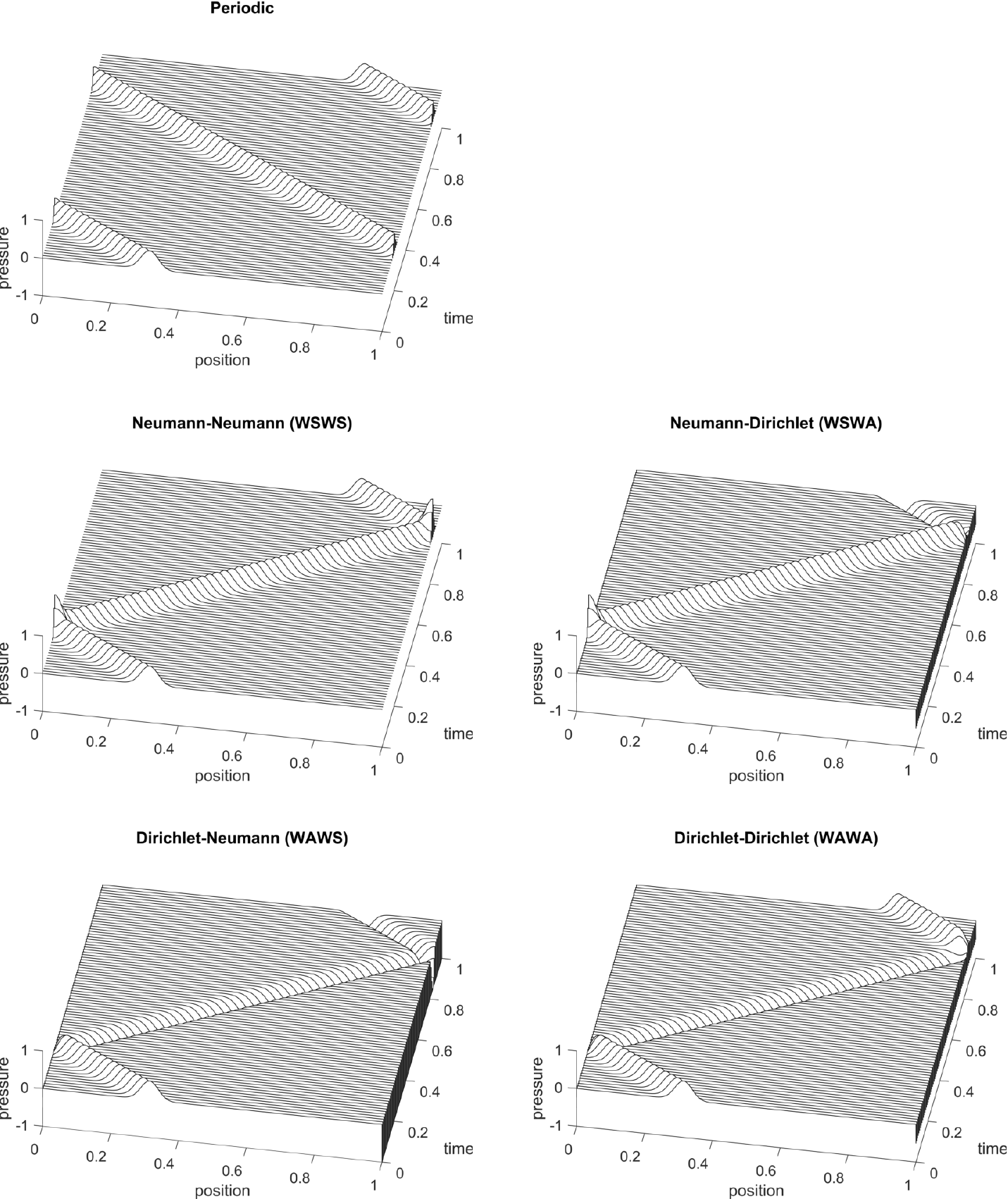}
    \caption{Propagation of a one-dimensional acoustic wave subject to different boundary conditions. The solution is calculated using a pseudospectral time domain method using either a Fourier basis or a sine or cosine basis for the spatial gradient calculation (see Table \ref{table_transform_choices}).}
    \label{fig_wave_eq_solution}
\end{figure}

The simulation results for each of the boundary conditions given in Table~\ref{table_transform_choices}  are shown in Fig.~\ref{fig_wave_eq_solution}. When a Fourier basis is used, the wave wraps from one side of the domain to the other when it reaches the boundary, a behaviour that is normally counteracted through the application of a perfectly matched layer. If a cosine or sine basis is used, the waves are instead reflected from the boundaries. When a  symmetric or Neumann boundary condition is used, the reflected wave has the same phase as the incident wave. When an antisymmetric or  Dirichlet boundary condition is used, the reflected wave is 180 degrees out-of-phase. 

To demonstrate the accuracy of the implemented boundary conditions, additional simulations were performed using a Neumman boundary condition at each end of the domain. In this case, the initial pressure distribution was set to a Gaussian in the centre of the grid and the initial particle velocity was set to zero (accounting for the staggered grid). The simulations were repeated for CFL numbers decreasing from $5\times10^{-1}$ to $5\times10^{-4}$ (the stability limit is $2/\pi$). For each simulation, the number of time steps was adjusted such that the outgoing waves propagated exactly to the boundaries and back to the centre of the grid where the reflected waves recombine to form the initial condition. The $L_\infty$ error was then calculated by comparing the calculated pressure field at the last time step with the initial condition.

\begin{figure}[!t]
    \centering
    \includegraphics[width=0.6\textwidth]{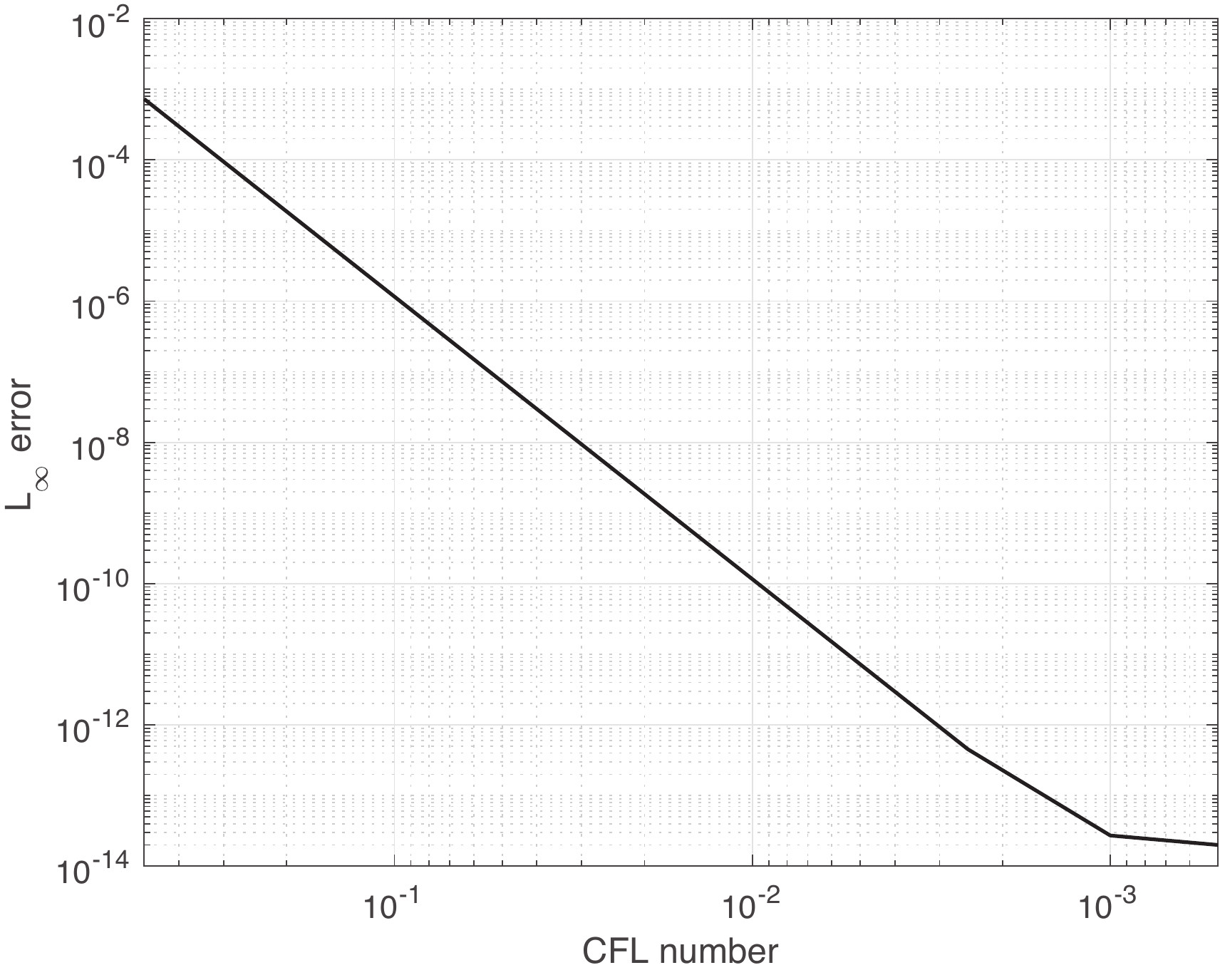}
    \caption{$L_\infty$ error in the pressure field after propagating a Gaussian initial pressure distribution from the centre of the computational domain to the boundaries and back using a DTT-based PSTD method with a Neumann boundary condition at each end of the domain.}
    \label{fig_convergence}
\end{figure}

The error convergence against CFL number is shown in Fig.\ \ref{fig_convergence}. In this case, the errors arise due to numerical dispersion introduced by the time stepping scheme, not due the accuracy of the boundary condition. As expected, the error reduces algebraically with the CFL number until machine precision is reached. The rate of convergence is governed by the order of the time stepping scheme used, as with conventional Fourier-based PSTD schemes \cite{boyd2001chebyshev}. For the chosen numerical scheme, engineering levels of accuracy, e.g., 0.1\%, can be reached with relatively modest CFL numbers.


\subsection{Higher dimensions}
\begin{figure}[!t]
    \includegraphics[width=\textwidth]{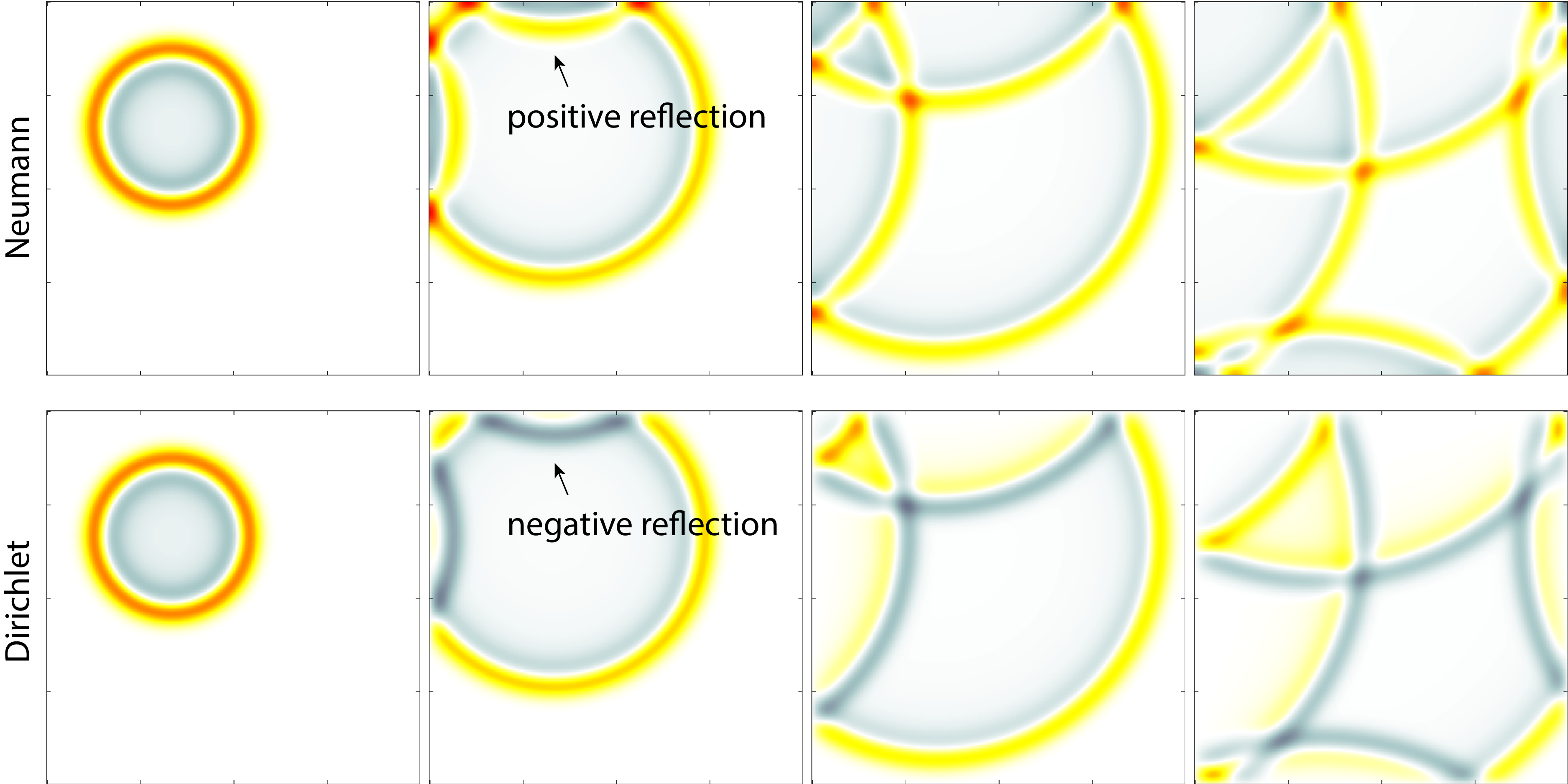}
    \caption{Four snapshots of the propagation of a two-dimensional acoustic wave in a rectangular cavity subject to Neumann and Dirichlet boundary conditions. The solution is calculated using a pseudospectral time domain method with a sine or cosine basis for the spatial gradient calculation (see Table \ref{table_transform_choices}).}
    \label{fig_wave_eq_2D}
\end{figure}

Much of the previous discussion focuses on the solution of the 1D wave equation using a DTT-based PSTD method. However, it is straightforward to extend this to higher dimensions. For example, in 2D, Eq.\ \eqref{eq_dtt_pstd_solution} becomes
\begin{align}
    u_x^{n+\tfrac{1}{2}} &= u_x^{n-\tfrac{1}{2}} - \frac{\Delta t}{\rho_0} \mathcal{T}_{x,2}^{-1}\left\{a_{x,1}\mathcal{T}_{x,1}\{p^n\}\right\} \enspace, \nonumber \\
    u_y^{n+\tfrac{1}{2}} &= u_y^{n-\tfrac{1}{2}} - \frac{\Delta t}{\rho_0} \mathcal{T}_{y,2}^{-1}\left\{a_{y,1}\mathcal{T}_{y,1}\{p^n\}\right\} \enspace, \nonumber \\ %
    p^{n+1} &= p^n - \Delta t \rho_0 c_0^2 \left( \mathcal{T}_{x,1}^{-1}\left\{a_{x,2}\mathcal{T}_{x,2}\left\{u_x^{n+\frac{1}{2}}\right\}\right\} + \mathcal{T}_{y,1}^{-1}\left\{a_{y,2}\mathcal{T}_{y,2}\left\{u_y^{n+\frac{1}{2}}\right\}\right\}\right) \enspace,
\end{align}
\noindent where $u_x$ and $u_y$ are the components of the particle velocity in the $x$ and $y$ directions. Again, the transforms $\mathcal{T}_1$ and $\mathcal{T}_2$ and {\kspace} coefficients $a_1$ and $a_2$ for the different boundary conditions are given in Table~\ref{table_transform_choices}. However, in this case, different boundary conditions can be chosen for the $x$ and $y$ directions (denoted by the subscript $x$ and $y$). This gives rise to 16 possible combinations of Neumann or Dirichlet boundary conditions on each of the four edges of the domain. Two examples are given in Fig.\ \ref{fig_wave_eq_2D} which shows the propagation of a Gaussian initial pressure distribution in a rectangular cavity using either Neumann or Dirichlet boundary conditions on all four boundaries. The application of DTT-based PSTD methods to the solution of other partial differential equations (for example, the diffusion equation) is also straightforward.

\subsection{Computational considerations}
In addition to the ability to implicitly enforce Dirichlet and Neumann boundary conditions, the use of a sine or cosine basis compared to a Fourier basis has several computational advantages. First, the DTTs are all real-to-real transforms (rather than real-to-complex or complex-to-complex). For 2D and 3D transforms, this means the output of the DTTs will consume half the memory required by the corresponding FFT (in 1D, if the input is real, the output of the FFT will be symmetric so only the unique half needs to be stored). Multiplications in {\kspace} are also cheaper to compute, as they operate on real data rather than complex. Finally, interpolation to and from staggered grids is done automatically by selecting the appropriate forward and inverse transforms, rather than by multiplying by a shift-factor in {\kspace}. 

In general, the computational complexity of computing the DCT and DST is $O(N\log N)$. However, according to Ref.\ \cite{frigo2018fftw}, due to implementation details, the transforms $\dct{2}$, $\dct{3}$, $\dst{2}$, $\dst{3}$ are the fastest, while the transforms $\dct{1}$, $\dst{1}$ can sometimes be significantly slower. The computation time is also significantly less when the implied period $M$ (sometimes called the logical DTT size) has small prime factors. This differs from the FFT, where the best performance is obtained when the sequence length $N$ has small prime factors. Table \ref{table_DTT_wavenumber_groups} outlines how to compute $M$ from $N$ for the different transform types.

Other properties of the DTT-based PSTD method are equivalent to conventional Fourier-based schemes. For example, in both cases, the wave-field is decomposed into a finite sum of trigonometric basis functions, and thus the same constraints on smoothness and convergence apply (see e.g., Ref.\ \cite{fornberg1998practical}). Similarly, the stability limits for the DTT-based scheme are identical to the Fourier case, and higher-order time stepping schemes can equally be used.


\section{Non-reflecting boundary conditions}
As described above and shown in Fig.~\ref{fig_wave_eq_solution}, DTTs can be used to model four basic pairs of boundary conditions in each dimension: Neumann-Neumann, Neumann-Dirichlet, Dirichlet-Neumann, and Dirichlet-Dirichlet. For convenience, these cases will be denoted in terms of the reflection coefficients of the left and right interfaces $(R_L,R_R)$: $(1,1)$, $(1,-1)$, $(-1,1)$, and $(-1,-1)$. Interestingly, the solution for more general boundary conditions can be constructed as a linear combination of these four cases (up to a time limit). In other words, the field that would result from any arbitrary real reflection coefficients on the left and right boundaries can be found by summing the fields from these four cases with suitably chosen weights.

\begin{figure}[!t]
    \centering
    \includegraphics[width=8cm]{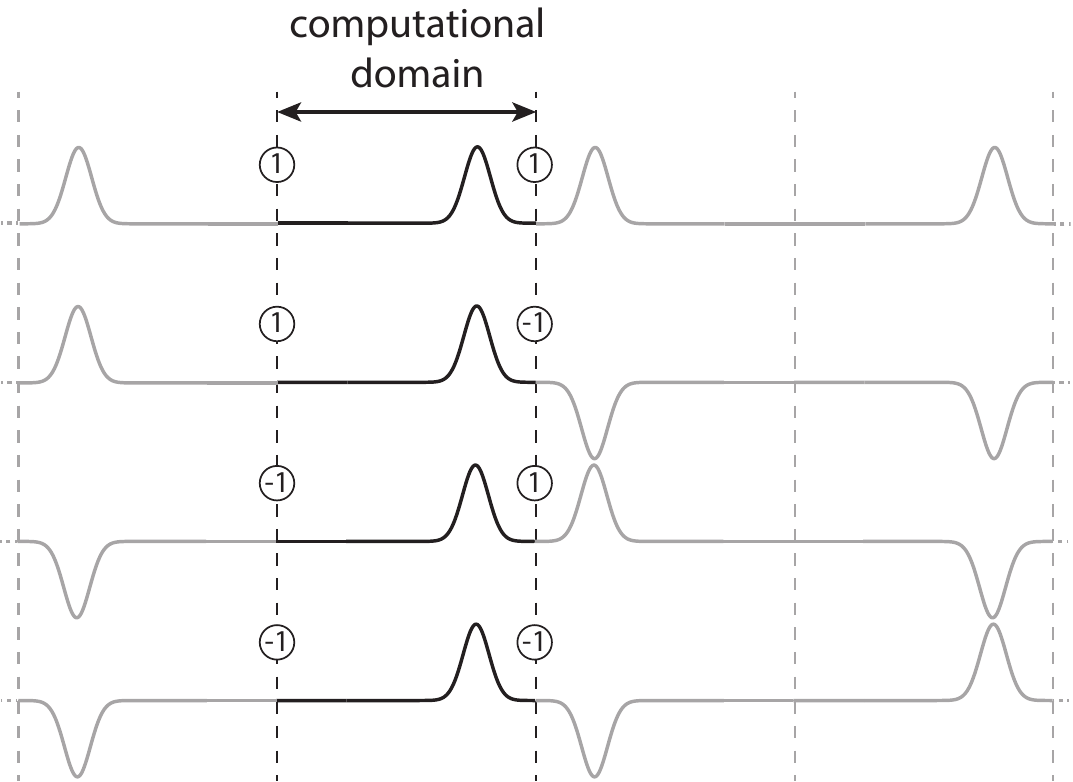}
    \caption{The implied image sources for the four basic cases of Neumann-Neumann, Neumann-Dirichlet, Dirichlet-Neumann, and Dirichlet-Dirichlet boundary conditions. The corresponding reflection coefficients are shown in circles. One periodic unit is shown. (Fig.~\ref{fig_implied_periodic_image_sources2} shows an extended case.)}
    \label{fig_implied_periodic_image_sources}
\end{figure}

To see how to choose the weights, first consider the implied periodicity in the four basic cases, as shown in Fig.~\ref{fig_implied_periodic_image_sources}.  In particular, notice the polarity and magnitude of the image sources. These can be written into a matrix, where each column corresponds to the polarity of the image sources for a particular pair of boundary conditions, i.e., each column in $\mathrm{M}$ corresponds to a row in Fig.~\ref{fig_implied_periodic_image_sources}:
\begin{equation}
\mathrm{M} = 
\left[
\begin{array}{rrrr}
1 &  1 & -1 & -1 \\
1 &  1 &  1 &  1 \\
1 & -1 &  1 & -1 \\
1 & -1 & -1 &  1
\end{array}
\right].
\end{equation}
Note that $\mathrm{M}/2$ is unitary. 
Now, if the four fields calculated using the different sets of boundary conditions are written as $\mathbf{p} = [p_{(1,1)}, p_{(1,-1)}, p_{(-1,1)}, p_{(-1,-1)}]^T$, and the corresponding weights for these fields as $\mathbf{w} = [w_{(1,1)}, w_{(1,-1)}, w_{(-1,1)}, w_{(-1,-1)}]^T$, then the field for an arbitrary reflection coefficient at each end of the domain can be written as
\begin{align}
p_{(R_L,R_R)}(x,t) = \mathbf{w}^T\mathbf{p}(x,t) \enspace.
\end{align}
The weights are calculated from
\begin{align}
\mathbf{w} = \tfrac{1}{4}\mathrm{M}^T\mathbf{r} \enspace,
\end{align}
where $\mathbf{r}$ corresponds to the amplitude of the implied image sources and is given by $\mathbf{r} = [R_L, 1, R_R, R_LR_R]^T$.

\begin{figure}[t!]
    \centering
    \includegraphics[width=\textwidth]{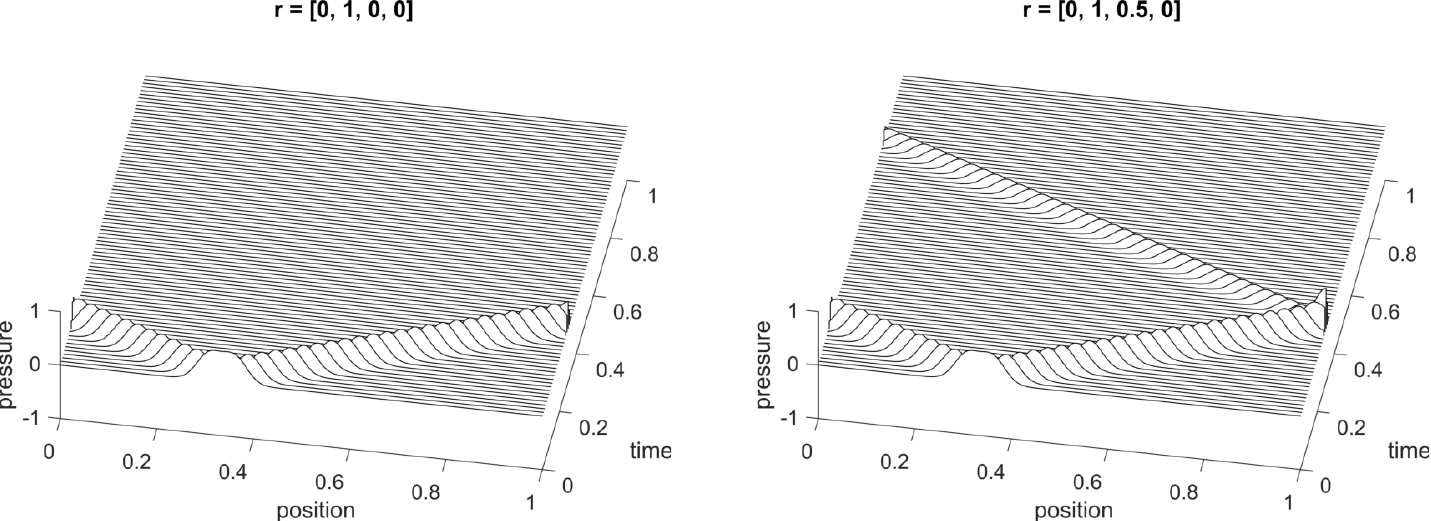}
    \caption{Propagation of a one-dimensional acoustic wave subject to different reflection coefficients at each end of the domain. The chosen reflection coefficients define the amplitudes of the implied image sources $\mathbf{r}$ and the weightings for the four basic fields $\mathbf{w}$. (left panel) $R_L=0$ and $R_R=0$. (right panel) $R_L=0$ and $R_R=0.5$.}
    \label{fig_wave_eq_non_reflecting}
\end{figure}

\begin{figure}[t!]
    \centering
    \includegraphics[width=\textwidth]{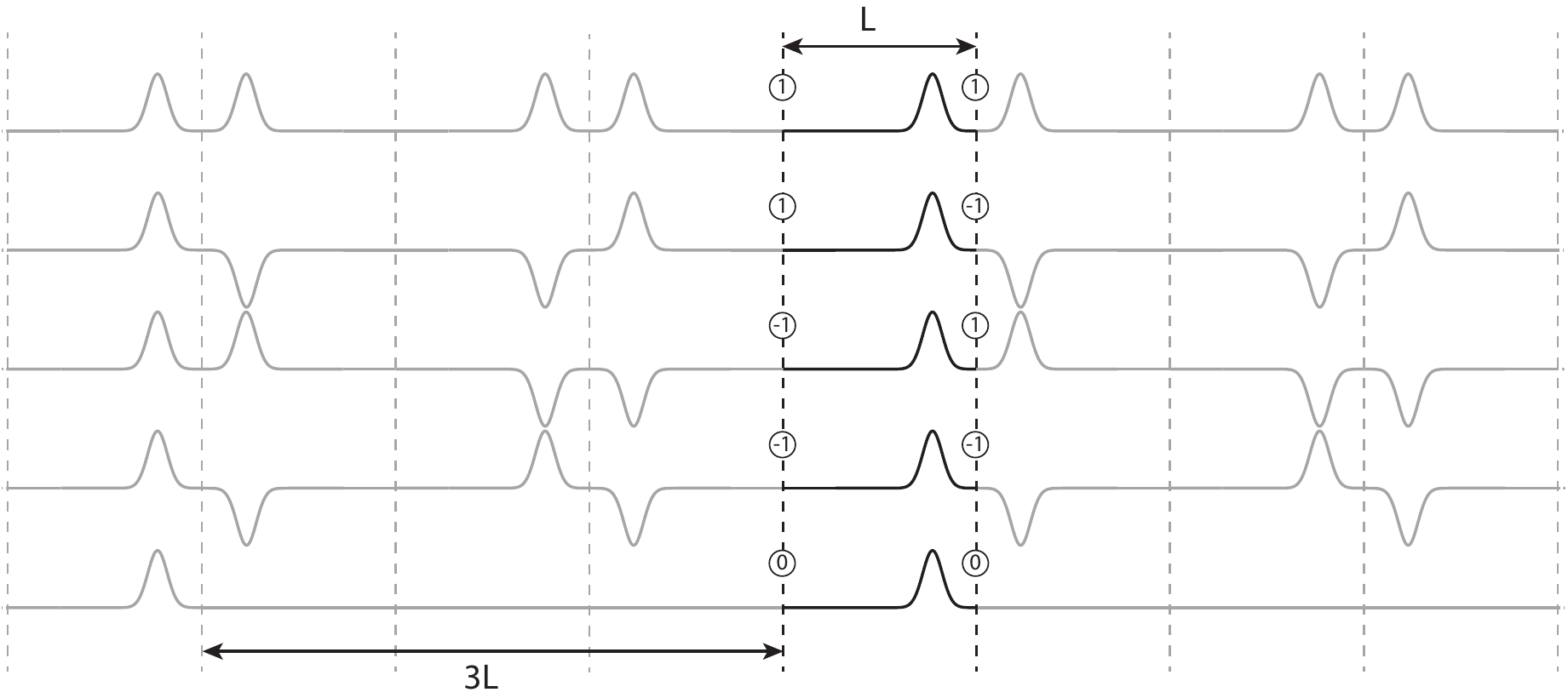}
    \caption{An extension of Fig.~\ref{fig_implied_periodic_image_sources}, with the bottom line showing the weighted summation of the four basic fields above. This represents the case where the reflection coefficients are both zero. The reflection coefficients are shown in circles. The reason for the time limit of $T=3L/c_0$ in this case is that the inherent periodicity in the transforms means that after this time there will be an arrival in the computational domain from the image source on the left.}
    \label{fig_implied_periodic_image_sources2}
\end{figure}

One example of interest is the case where both boundaries are non-reflecting, i.e., $R_L = R_R = 0$. It can immediately be seen that the weights in this case are
\begin{equation}
\left[
\begin{array}{c}
w_{(1,1)} \\
w_{(1,-1)} \\
w_{(-1,1)} \\
w_{(-1,-1)}
\end{array}
\right]
=
\frac{1}{4}
\left[
\begin{array}{rrrr}
 1 & \hphantom{-}1 &  1 &  1 \\
 1 & 1 & -1 & -1 \\
-1 & 1 &  1 & -1 \\
-1 & 1 & -1 &  1
\end{array}
\right]
\left[
\begin{array}{c}
0 \\
1 \\
0 \\
0
\end{array}
\right]
=
\frac{1}{4}
\left[
\begin{array}{c}
1 \\
1 \\
1 \\
1
\end{array}
\right].
\end{equation}
In other words, the field in the case of non-reflecting boundaries can be constructed by adding together the four basic fields and dividing by four (see Fig.~\ref{fig_implied_periodic_image_sources2}). This case is discussed in detail in Ref.\ \cite{smith1974nonreflecting}. Other cases can be calculated similarly. For example, the case for $R_L=0$ and $R_R=0.5$ requires weights $\mathbf{w}=[3, 1, 3, 1]^T/8$. Simulation results for these two cases are given in Fig.\ \ref{fig_wave_eq_non_reflecting}. In this example, the initial pressure is defined as a Gaussian and the initial particle velocity is set to zero, and thus the wave propagates in both directions. After the waves have left the domain, the wave cancellation is theoretically exact, and for the examples give in Fig.\ \ref{fig_wave_eq_non_reflecting} the pressure is zero to machine precision.

While this approach allows boundaries with arbitrary real reflection coefficients to be defined (including non-reflecting boundaries), there are two important limitations. First, the linear combination approach only works for times $t = [0,T)$. At time $T$, additional waves enter the field. The reason for this limit is the inherent periodicity, which is visualised for the non-reflecting case in Fig.~\ref{fig_implied_periodic_image_sources2}. In the case when both boundaries are non-reflecting, the limit is $T = 3L/c_0$, where $L$ is the size of the computational domain, although in the general case the limit is stricter, $T = 2L/c_0$. Second, the calculation requires four models to be run in the 1D case, although the models are independent and so can be computed in parallel. The same idea (with the same time limits) can be extended to higher dimensions, however, the number of basic models required grows as $2^{2d}$, where $d$ is the number of spatial dimensions. Thus to form a non-reflecting boundary requires 16 simulations in 2D, and 64 in 3D. On the other hand, within the time limits mentioned above, the approach is theoretically exact, which may be an advantage in some situations over conventional absorbing boundary conditions or perfectly matched layers which typically only give a few decimal points of accuracy.

\section{Summary}
A DTT-based PSTD method for the wave equation is described that allows Neumann (sound-hard) and Dirichlet (sound-soft) boundary conditions to be imposed. The spatial gradients are calculated using a spectral collocation method with a discrete cosine or sine basis, and time integration is performed using a conventional finite difference scheme. The different combinations of staggered grids and boundary conditions lead to the use of sixteen different discrete trigonometric transforms (8 DCTs and 8 DTSs). These can be grouped into four distinct classes that are closed under differentiation and staggered-grid interpolation. Practical details of how to use these transforms to compute spatial gradients are provided. Numerical examples of the solution to the wave equation in 1D and 2D in a rectangular domain subject to different boundary conditions are also given. Finally, the extension to boundaries with arbitrary real reflection coefficients is discussed. The DTT library and example scripts used in this paper are available open-source from GitHub.\cite{treeby2020dttlibrary}

\section*{Acknowledgment}
This work was supported by the Engineering and Physical Sciences Research Council (EPSRC), UK, under grant numbers EP/M011119/1, EP/L020262/1, EP/P008860/1, and EP/S026371/1. This work was also supported by The Ministry of Education, Youth and Sports of the Czech Republic from the National Programme of Sustainability (NPU II); project IT4Innovations excellence in science - LQ1602. This work was also supported in part by the Australian Research Council/Microsoft Linkage Project LP100100588.


\bibliographystyle{ieeetr}


\section*{Appendix: Tables describing DTT methods}
\begin{table}[h!]
\renewcommand{\arraystretch}{1.4}
\centering
\caption{Differentiation for periodic SPSs.}
\resizebox{\textwidth}{!}{%
\begin{tabular}{c l l l l c}
\hline
\multicolumn{3}{c}{Original sequence $\vec{u}$} & \multicolumn{3}{c}{Final sequence $\vec{v}$} \\
Symmetry &	Representative sample & Forward & {\kspace} operation & Inverse & Symmetry \\
\hline
WSWS  &
$\vec{u}=\{u_0,~\ldots,~u_\frac{M}{2}\}$  &
$\hat{\vec{u}}=\dct{1}\vec{u}$ &
$\hat{\vec{v}} = \{-k_1\hat{u}_1,~\ldots,~-k_{\frac{M}{2}-1}\hat{u}_{\frac{M}{2}-1}\}$ & 
$\vec{v} = \frac{1}{M}\dst{1}\hat{\vec{v}}$ & WAWA \\
HSHS &
$\vec{u}=\{u_0,\ldots,u_{\frac{M}{2}-1}\}$ &
$\hat{\vec{u}}=\dct{2}\vec{u}$ &
$\hat{\vec{v}} = \{-k_1\hat{u}_1,~\ldots,-k_{\frac{M}{2}-1}\hat{u}_{\frac{M}{2}-1},~0\}$ & 
$\vec{v}=\frac{1}{M}\dst{3}\hat{\vec{v}}$ & HAHA \\
WAWA &
$\vec{u}=\{u_1,~\ldots,~u_{\frac{M}{2}-1}\}$ &
$\hat{\vec{u}}=\dst{1}\vec{u}$ &
$\hat{\vec{v}}=\{0,~k_1\hat{u}_1,~\ldots,~k_{\frac{M}{2}-1}\hat{u}_{\frac{M}{2}-1},0\}$ & 
$\vec{v} = \frac{1}{M}\dct{1}\hat{\vec{v}}$ & WSWS \\
HAHA &
$\vec{u}=\{u_1,~\ldots,u_\frac{M}{2}\}$ &
$\hat{\vec{u}}=\dst{2}\vec{u}$ &
$\hat{\vec{v}}=\{0, k_1\hat{u}_1,\ldots,k_{\frac{M}{2}-1}\hat{u}_{\frac{M}{2}-1}\}$ & 
$\vec{v}=\frac{1}{M}\dct{3}\hat{\vec{v}}$ & HSHS \\
\hline
WSHS &
$\vec{u}=\{u_0,~\ldots,u_\frac{M-1}{2}\}$ &
$\hat{\vec{u}}=\dct{5}\vec{u}$ &
$\hat{\vec{v}}=\{-k_1\hat{u}_1,\ldots,-k_\frac{M-1}{2}\hat{u}_\frac{M-1}{2},~0\}$ & 
$\vec{v}=\frac{1}{M}\dst{5}\hat{\vec{v}}$ & WAHA \\
HSWS &
$\vec{u}=\{u_0,~\ldots,u_\frac{M-1}{2}\}$ &
$\hat{\vec{u}}=\dct{6}\vec{u}$ &
$\hat{\vec{v}}=\{-k_1\hat{u}_1,\ldots,-k_\frac{M-1}{2}\hat{u}_\frac{M-1}{2},~0\}$ & 
$\vec{v}=\frac{1}{M}\dst{7}\hat{\vec{v}}$ & HAWA \\
WAHA &
$\vec{u}=\{u_1,~\ldots,u_\frac{M+1}{2}\}$ &
$\hat{\vec{u}}=\dst{5}\vec{u}$ &
$\hat{\vec{v}}=\{0,~k_1\hat{u}_1,\ldots,k_\frac{M-1}{2}\hat{u}_\frac{M-1}{2}\}$ & 
$\vec{v}=\frac{1}{M}\dct{5}\hat{\vec{v}}$ & WSHS \\
HAWA &
$\vec{u}=\{u_1,~\ldots,u_\frac{M+1}{2}\}$ &
$\hat{\vec{u}}=\dst{6}\vec{u}$ &
$\hat{\vec{v}}=\{0,~k_1\hat{u}_1,\ldots,k_\frac{M-1}{2}\hat{u}_\frac{M-1}{2}\}$ & 
$\vec{v}=\frac{1}{M}\dct{7}\hat{\vec{v}}$ & HSWS \\
\hline
\end{tabular}}
\label{table_diff_periodic_sps}
\end{table}
\begin{table}[h!]
\renewcommand{\arraystretch}{1.4}
\centering
\caption{Interpolation between staggered grids for periodic SPSs.}
\resizebox{\textwidth}{!}{%
\begin{tabular}{c l l l l c}
\hline
\multicolumn{3}{c}{Original sequence $\vec{u}$} & \multicolumn{3}{c}{Final sequence $\vec{v}$} \\
Symmetry &	Representative sample & Forward & {\kspace} operation & Inverse & Symmetry \\
\hline
WSWS  &
$\vec{u}=\{u_0,~\ldots,~u_\frac{M}{2}\}$  &
$\hat{\vec{u}}=\dct{1}\vec{u}$ &
$\hat{\vec{v}} = \{\hat{u}_0,~\ldots,~\hat{u}_{\frac{M}{2}-1}\}$ & 
$\vec{v} = \frac{1}{M}\dct{3}\hat{\vec{v}}$ & HSHS \\
HSHS &
$\vec{u}=\{u_0,\ldots,u_{\frac{M}{2}-1}\}$ &
$\hat{\vec{u}}=\dct{2}\vec{u}$ &
$\hat{\vec{v}} = \{\hat{u}_1,~\ldots,\hat{u}_{\frac{M}{2}-1},~0\}$ & 
$\vec{v}=\frac{1}{M}\dct{1}\hat{\vec{v}}$ & WSWS \\
WAWA &
$\vec{u}=\{u_1,~\ldots,~u_{\frac{M}{2}-1}\}$ &
$\hat{\vec{u}}=\dst{1}\vec{u}$ &
$\hat{\vec{v}}=\{\hat{u}_1,~\ldots,~\hat{u}_{\frac{M}{2}-1},~0\}$ & 
$\vec{v} = \frac{1}{M}\dst{3}\hat{\vec{v}}$ & HAHA \\
HAHA &
$\vec{u}=\{u_1,~\ldots,u_\frac{M}{2}\}$ &
$\hat{\vec{u}}=\dst{2}\vec{u}$ &
$\hat{\vec{v}}=\{\hat{u}_1,\ldots,\hat{u}_{\frac{M}{2}-1}\}$ & 
$\vec{v}=\frac{1}{M}\dst{1}\hat{\vec{v}}$ & WAWA \\
\hline
WSHS &
$\vec{u}=\{u_0,~\ldots,u_\frac{M-1}{2}\}$ &
$\hat{\vec{u}}=\dct{5}\vec{u}$ &
$\hat{\vec{v}}=\hat{\vec{u}}$ & 
$\vec{v}=\frac{1}{M}\dct{7}\hat{\vec{v}}$ & HSWS \\
HSWS &
$\vec{u}=\{u_0,~\ldots,u_\frac{M-1}{2}\}$ &
$\hat{\vec{u}}=\dct{6}\vec{u}$ &
$\hat{\vec{v}}=\hat{\vec{u}}$ & 
$\vec{v}=\frac{1}{M}\dct{5}\hat{\vec{v}}$ & WSHS \\
WAHA &
$\vec{u}=\{u_1,~\ldots,u_\frac{M+1}{2}\}$ &
$\hat{\vec{u}}=\dst{5}\vec{u}$ &
$\hat{\vec{v}}=\hat{\vec{u}}$ & 
$\vec{v}=\frac{1}{M}\dst{7}\hat{\vec{v}}$ & HAWA \\
HAWA &
$\vec{u}=\{u_1,~\ldots,u_\frac{M+1}{2}\}$ &
$\hat{\vec{u}}=\dst{6}\vec{u}$ &
$\hat{\vec{v}}=\hat{\vec{u}}$ & 
$\vec{v}=\frac{1}{M}\dst{5}\hat{\vec{v}}$ & WAHA \\
\hline
\end{tabular}}
\label{table_interp_periodic_sps}
\end{table}
\begin{table}[h!]
\renewcommand{\arraystretch}{1.4}
\centering
\caption{Differentiation with staggered grids for periodic SPSs.}
\resizebox{\textwidth}{!}{%
\begin{tabular}{c l l l l c}
\hline
\multicolumn{3}{c}{Original sequence $\vec{u}$} & \multicolumn{3}{c}{Final sequence $\vec{v}$} \\
Symmetry &	Representative sample & Forward & {\kspace} operation & Inverse & Symmetry \\
\hline
WSWS  &
$\vec{u}=\{u_0,~\ldots,~u_\frac{M}{2}\}$  &
$\hat{\vec{u}}=\dct{1}\vec{u}$ &
$\hat{\vec{v}} = \{-k_1\hat{u}_1,~\ldots,~-k_\frac{M}{2}\hat{u}_\frac{M}{2}\}$ & 
$\vec{v} = \frac{1}{M}\dst{3}\hat{\vec{v}}$ & HAHA \\
HSHS &
$\vec{u}=\{u_0,\ldots,u_{\frac{M}{2}-1}\}$ &
$\hat{\vec{u}}=\dct{2}\vec{u}$ &
$\hat{\vec{v}} = \{-k_1\hat{u}_1,~\ldots,-k_{\frac{M}{2}-1}\hat{u}_{\frac{M}{2}-1}\}$ & 
$\vec{v}=\frac{1}{M}\dst{1}\hat{\vec{v}}$ & WAWA \\
WAWA &
$\vec{u}=\{u_1,~\ldots,~u_{\frac{M}{2}-1}\}$ &
$\hat{\vec{u}}=\dst{1}\vec{u}$ &
$\hat{\vec{v}}=\{0,~k_1\hat{u}_1,~\ldots,~k_{\frac{M}{2}-1}\hat{u}_{\frac{M}{2}-1}\}$ & 
$\vec{v} = \frac{1}{M}\dct{3}\hat{\vec{v}}$ & HSHS \\
HAHA &
$\vec{u}=\{u_1,~\ldots,u_\frac{M}{2}\}$ &
$\hat{\vec{u}}=\dst{2}\vec{u}$ &
$\hat{\vec{v}}=\{0, k_1\hat{u}_1,\ldots,k_\frac{M}{2}\hat{u}_\frac{M}{2}\}$ & 
$\vec{v}=\frac{1}{M}\dct{1}\hat{\vec{v}}$ & WSWS \\
\hline
WSHS &
$\vec{u}=\{u_0,~\ldots,u_\frac{M-1}{2}\}$ &
$\hat{\vec{u}}=\dct{5}\vec{u}$ &
$\hat{\vec{v}}=\{-k_1\hat{u}_1,\ldots,-k_\frac{M-1}{2}\hat{u}_\frac{M-1}{2},~0\}$ & 
$\vec{v}=\frac{1}{M}\dst{7}\hat{\vec{v}}$ & HAWA \\
HSWS &
$\vec{u}=\{u_0,~\ldots,u_\frac{M-1}{2}\}$ &
$\hat{\vec{u}}=\dct{6}\vec{u}$ &
$\hat{\vec{v}}=\{-k_1\hat{u}_1,\ldots,-k_\frac{M-1}{2}\hat{u}_\frac{M-1}{2},~0\}$ & 
$\vec{v}=\frac{1}{M}\dst{5}\hat{\vec{v}}$ & WAHA \\
WAHA &
$\vec{u}=\{u_1,~\ldots,u_\frac{M+1}{2}\}$ &
$\hat{\vec{u}}=\dst{5}\vec{u}$ &
$\hat{\vec{v}}=\{0,~k_1\hat{u}_1,\ldots,k_\frac{M-1}{2}\hat{u}_\frac{M-1}{2}\}$ & 
$\vec{v}=\frac{1}{M}\dct{7}\hat{\vec{v}}$ & HSWS \\
HAWA &
$\vec{u}=\{u_1,~\ldots,u_\frac{M+1}{2}\}$ &
$\hat{\vec{u}}=\dst{6}\vec{u}$ &
$\hat{\vec{v}}=\{0,~k_1\hat{u}_1,\ldots,k_\frac{M-1}{2}\hat{u}_\frac{M-1}{2}\}$ & 
$\vec{v}=\frac{1}{M}\dct{5}\hat{\vec{v}}$ & WSHS \\
\hline
\end{tabular}}
\label{table_diff_interp_periodic_sps}
\end{table}
\begin{table}[h!]
\renewcommand{\arraystretch}{1.4}
\centering
\caption{Differentiation for antiperiodic SPSs.}
\resizebox{\textwidth}{!}{%
\begin{tabular}{c l l l l c}
\hline
\multicolumn{3}{c}{Original sequence $\vec{u}$} & \multicolumn{3}{c}{Final sequence $\vec{v}$} \\
Symmetry &	Representative sample & Forward & {\kspace} operation & Inverse & Symmetry \\
\hline
WSWA & 
$\vec{u}=\{u_0,~\ldots,~u_{\frac{M}{2}-1}\}$  &
$\hat{\vec{u}}=\dct{3}\vec{u}$ &
$\hat{\vec{v}}=-\vec{k}\cdot\hat{\vec{u}}$ & 
$\vec{v}=\frac{1}{M}\dst{2}\hat{\vec{v}}$ & WAWS \\
HSHA & 
$\vec{u}=\{u_0,~\ldots,~u_{\frac{M}{2}-1}\}$  &
$\hat{\vec{u}}=\dct{4}\vec{u}$ &
$\hat{\vec{v}}=-\vec{k}\cdot\hat{\vec{u}}$ & 
$\vec{v}=\frac{1}{M}\dst{4}\hat{\vec{v}}$ & HAHS \\
WAWS & 
$\vec{u}=\{u_0,~\ldots,~u_{\frac{M}{2}-1}\}$  &
$ \hat{\vec{u}}=\dst{3}\vec{u}$ & 
$\hat{\vec{v}}=\vec{k}\cdot\hat{\vec{u}}$ & 
$\vec{v}=\frac{1}{M}\dct{2}\hat{\vec{v}}$ & WSWA \\
HAHS & 
$\vec{u}=\{u_0,~\ldots,~u_{\frac{M}{2}-1}\}$  &
$\hat{\vec{u}}=\dst{4}\vec{u}$ & 
$\hat{\vec{v}}=\vec{k}\cdot\hat{\vec{u}}$ & 
$\vec{v}=\frac{1}{M}\dct{4}\hat{\vec{v}}$ & HSHA \\
\hline
WSHA & 
$\vec{u}=\{u_0,~\ldots,~u_\frac{M-1}{2}\}$  &
$\hat{\vec{u}}=\dct{7}\vec{u}$ &
$\hat{\vec{v}}=-\vec{k}\cdot\hat{\vec{u}}$ & 
$\vec{v}=\frac{1}{M}\dst{6}\hat{\vec{v}}$ & WAHS \\
HSWA & 
$\vec{u}=\{u_0,~\ldots,~u_\frac{M-1}{2}\}$  &
$\hat{\vec{u}}=\dct{8}\vec{u}$ &
$\hat{\vec{v}}=-\vec{k}\cdot\hat{\vec{u}}$ & 
$\vec{v}=\frac{1}{M}\dst{8}\hat{\vec{v}}$ & HAWS \\
WAHS & 
$\vec{u}=\{u_0,~\ldots,~u_\frac{M-1}{2}\}$  &
$\hat{\vec{u}}=\dst{7}\vec{u}$ & 
$\hat{\vec{v}}=\vec{k}\cdot\hat{\vec{u}}$ & 
$\vec{v}=\frac{1}{M}\dct{6}\hat{\vec{v}}$ & WSHA \\
HAWS & 
$\vec{u}=\{u_0,~\ldots,~u_\frac{M-1}{2}\}$  &
$\hat{\vec{u}}=\dst{8}\vec{u}$ & 
$\hat{\vec{v}}=\vec{k}\cdot\hat{\vec{u}}$ & 
$\vec{v}=\frac{1}{M}\dct{8}\hat{\vec{v}}$ & HSWA \\
\hline
\end{tabular}}
\label{table_diff_antiperiodic_sps}
\end{table}
\begin{table}[h!]
\renewcommand{\arraystretch}{1.4}
\centering
\caption{Interpolation between staggered grids for antiperiodic SPSs.}
\resizebox{\textwidth}{!}{%
\begin{tabular}{c l l l l c}
\hline
\multicolumn{3}{c}{Original sequence $\vec{u}$} & \multicolumn{3}{c}{Final sequence $\vec{v}$} \\
Symmetry &	Representative sample & Forward & {\kspace} operation & Inverse & Symmetry \\
\hline
WSWA & 
$\vec{u}=\{u_0,~\ldots,~u_{\frac{M}{2}-1}\}$  &
$\hat{\vec{u}}=\dct{3}\vec{u}$ & 
$\hat{\vec{v}}=\hat{\vec{u}}$ & 
$\vec{v}=\frac{1}{M}\dct{4}\hat{\vec{v}}$ & HSHA \\
HSHA & 
$\vec{u}=\{u_0,~\ldots,~u_{\frac{M}{2}-1}\}$  &
$\hat{\vec{u}}=\dct{4}\vec{u}$ & 
$\hat{\vec{v}}=\hat{\vec{u}}$ &  
$\vec{v}=\frac{1}{M}\dct{2}\hat{\vec{v}}$ & WSWA \\
WAWS & 
$\vec{u}=\{u_0,~\ldots,~u_{\frac{M}{2}-1}\}$  &
$\hat{\vec{u}}=\dst{4}\vec{u}$ & 
$\hat{\vec{v}}=\hat{\vec{u}}$ &  
$\vec{v}=\frac{1}{M}\dst{4}\hat{\vec{v}}$ & HAHS \\
HAHS & 
$\vec{u}=\{u_0,~\ldots,~u_{\frac{M}{2}-1}\}$  &
$\hat{\vec{u}}=\dst{4}\vec{u}$ & 
$\hat{\vec{v}}=\hat{\vec{u}}$ &  
$\vec{v}=\frac{1}{M}\dst{2}\hat{\vec{v}}$ & WAWS \\
\hline
WSHA & 
$\vec{u}=\{u_0,~\ldots,~u_\frac{M-1}{2}\}$  &
$\hat{\vec{u}}=\dct{7}\vec{u}$ & 
$\hat{\vec{v}}=\hat{\vec{u}}$ &  
$\vec{v}=\frac{1}{M}\dct{8}\hat{\vec{v}}$ & HSWA \\
HSWA & 
$\vec{u}=\{u_0,~\ldots,~u_\frac{M-1}{2}\}$  &
$\hat{\vec{u}}=\dct{8}\vec{u}$ & 
$\hat{\vec{v}}=\hat{\vec{u}}$ &  
$\vec{v}=\frac{1}{M}\dct{6}\hat{\vec{v}}$ & WSHA \\
WAHS & 
$\vec{u}=\{u_0,~\ldots,~u_\frac{M-1}{2}\}$  &
$\hat{\vec{u}}=\dst{7}\vec{u}$ & 
$\hat{\vec{v}}=\hat{\vec{u}}$ &  
$\vec{v}=\frac{1}{M}\dst{8}\hat{\vec{v}}$ & HAWS \\
HAWS & 
$\vec{u}=\{u_0,~\ldots,~u_\frac{M-1}{2}\}$  &
$\hat{\vec{u}}=\dst{8}\vec{u}$ & 
$\hat{\vec{v}}=\hat{\vec{u}}$ &  
$\vec{v}=\frac{1}{M}\dst{6}\hat{\vec{v}}$ & WAHS \\
\hline
\end{tabular}}
\label{table_interp_antiperiodic_sps}
\end{table}
\begin{table}[h!]
\renewcommand{\arraystretch}{1.4}
\centering
\caption{Differentiation with staggered grids for antiperiodic SPSs.}
\resizebox{\textwidth}{!}{%
\begin{tabular}{c l l l l c}
\hline
\multicolumn{3}{c}{Original sequence $\vec{u}$} & \multicolumn{3}{c}{Final sequence $\vec{v}$} \\
Symmetry &	Representative sample & Forward & {\kspace} operation & Inverse & Symmetry \\
\hline
WSWA & 
$\vec{u}=\{u_0,~\ldots,~u_{\frac{M}{2}-1}\}$  &
$\hat{\vec{u}}=\dct{3}\vec{u}$ & 
$\hat{\vec{v}}=-\vec{k}\cdot\hat{\vec{u}}$ & 
$\vec{v}=\frac{1}{M}\dst{4}\hat{\vec{v}}$ & HAHS \\
HSHA & 
$\vec{u}=\{u_0,~\ldots,~u_{\frac{M}{2}-1}\}$  &
$\hat{\vec{u}}=\dct{4}\vec{u}$ & 
$\hat{\vec{v}}=-\vec{k}\cdot\hat{\vec{u}}$ & 
$\vec{v}=\frac{1}{M}\dst{2}\hat{\vec{v}}$ & WAWS \\
WAWS & 
$\vec{u}=\{u_0,~\ldots,~u_{\frac{M}{2}-1}\}$  &
$\hat{\vec{u}}=\dst{4}\vec{u}$ & 
$\hat{\vec{v}}=\vec{k}\cdot\hat{\vec{u}}$ & 
$\vec{v}=\frac{1}{M}\dct{4}\hat{\vec{v}}$ & HSHA \\
HAHS & 
$\vec{u}=\{u_0,~\ldots,~u_{\frac{M}{2}-1}\}$  &
$\hat{\vec{u}}=\dst{4}\vec{u}$ & 
$\hat{\vec{v}}=\vec{k}\cdot\hat{\vec{u}}$ & 
$\vec{v}=\frac{1}{M}\dct{2}\hat{\vec{v}}$ & WSWA \\
\hline
WSHA & 
$\vec{u}=\{u_0,~\ldots,~u_\frac{M-1}{2}\}$  &
$\hat{\vec{u}}=\dct{7}\vec{u}$ & 
$\hat{\vec{v}}=-\vec{k}\cdot\hat{\vec{u}}$ & 
$\vec{v}=\frac{1}{M}\dst{8}\hat{\vec{v}}$ & HAWS \\
HSWA & 
$\vec{u}=\{u_0,~\ldots,~u_\frac{M-1}{2}\}$  &
$\hat{\vec{u}}=\dct{8}\vec{u}$ & 
$\hat{\vec{v}}=-\vec{k}\cdot\hat{\vec{u}}$ & 
$\vec{v}=\frac{1}{M}\dst{6}\hat{\vec{v}}$ & WAHS \\
WAHS & 
$\vec{u}=\{u_0,~\ldots,~u_\frac{M-1}{2}\}$  &
$\hat{\vec{u}}=\dst{7}\vec{u}$ & 
$\hat{\vec{v}}=\vec{k}\cdot\hat{\vec{u}}$ & 
$\vec{v}=\frac{1}{M}\dct{8}\hat{\vec{v}}$ & HSWA \\
HAWS & 
$\vec{u}=\{u_0,~\ldots,~u_\frac{M-1}{2}\}$  &
$\hat{\vec{u}}=\dst{8}\vec{u}$ & 
$\hat{\vec{v}}=\vec{k}\cdot\hat{\vec{u}}$ & 
$\vec{v}=\frac{1}{M}\dct{6}\hat{\vec{v}}$ & WSHA \\
\hline
\end{tabular}}
\label{table_diff_interp_antiperiodic_sps}
\end{table}

\end{document}